\documentclass[aps,prx,reprint,preprintnumbers,superscriptaddress,nofootinbib,longbibliography,floatfix]{revtex4-2}
\pdfoutput=1
\usepackage{rotating}
\usepackage{array}
\usepackage{amsmath}
\usepackage[normalem]{ulem}
\usepackage{slashed}
\usepackage{booktabs}
\usepackage[pdftex,table]{xcolor}
\usepackage{units}
\usepackage{xfrac}
\usepackage{xspace}
\usepackage{mathtools}
\usepackage{empheq}
\usepackage[]{units}
\usepackage{multirow}
\usepackage{amssymb}
\usepackage{url}
\usepackage{comment}
\usepackage{physics}
\usepackage{color,soul}
\usepackage{bbm}
\usepackage[caption=false]{subfig}
\usepackage{adjustbox}

\usepackage{hyperref}
\hypersetup{
  colorlinks=true,
  citecolor=blue,
  linkcolor=blue,
  urlcolor=blue
}

\newcommand{\geant}{\textsc{Geant4}\xspace}
\newcommand{\gf}{\textsc{GFlash}\xspace}

\begin{document}

\title{Refining Fast Calorimeter Simulations with a Schr\"{o}dinger Bridge}

\author{Sascha Diefenbacher}
\email{sdiefenbacher@lbl.gov}
\affiliation{Physics Division, Lawrence Berkeley National Laboratory, Berkeley, CA 94720, USA}

\author{Vinicius Mikuni}
\email{vmikuni@lbl.gov}
\affiliation{National Energy Research Scientific Computing Center, Berkeley Lab, Berkeley, CA 94720, USA}

\author{Benjamin Nachman}
\email{bpnachman@lbl.gov}
\affiliation{Physics Division, Lawrence Berkeley National Laboratory, Berkeley, CA 94720, USA}
\affiliation{Berkeley Institute for Data Science, University of California, Berkeley, CA 94720, USA}

\begin{abstract}
Machine learning-based simulations, especially calorimeter simulations, are promising tools for approximating the precision of classical high energy physics simulations with a fraction of the generation time.  Nearly all methods proposed so far learn neural networks that map a random variable with a known probability density, like a Gaussian, to realistic-looking events.  In many cases, physics events are not close to Gaussian and so these neural networks have to learn a highly complex function.  We study an alternative approach: Schr\"{o}dinger bridge Quality Improvement via Refinement of Existing Lightweight Simulations (SQuIRELS). SQuIRELS leverages the power of diffusion-based neural networks and Schr\"{o}dinger bridges to map between samples where the probability density is not known explicitly.  We apply SQuIRELS to the task of refining a classical fast simulation to approximate a full classical simulation.  On simulated calorimeter events, we find that SQuIRELS is able to reproduce highly non-trivial features of the full simulation with a fraction of the generation time.

\end{abstract}

\maketitle


\section{Introduction}
\label{sec:intro}

Monte Carlo simulations are an essential tool for inference in high energy physics by connecting theory and experiment.  These simulations describe the complete evolution from fundamental interactions to detector effects.  Monte Carlo datasets must significantly exceed the size of experimental datasets in order to ensure that their contribution to the statistical uncertainty is insignificant.  For the most precise physics-based simulations, the speed of state-of-the-art generators can make this a computing challenge.  Calorimeters are often the slowest component of a complete simulation because they need to model the evolution of particles from their highest energies down to essentially no energy as they are stopped by the detector material.  Tracking all of the corresponding secondary particles inside the material using tools like \geant~\cite{geant4} can be prohibitively slow.

Well-established experiments like ATLAS, CMS, and LHCb at the Large Hadron Collider can mitigate the computational burden of physics-based simulations by using fast approximate (surrogate) models.  Classical fast simulations are built on parameterizations of the detector response and by modeling various components of detector effects as factorized from each other. While some components may be optimized through fits to a \geant-based (`full') simulation, most of the structure is engineered manually.  The resulting simulations are 10-100 times faster than \geant-based simulations and have been widely used when precise modeling is not required~\cite{ATLAS:2021pzo,Giammanco:2014bza,Barbetti:2023bvi}.  As precision requires the use of lower-level features combined with deep learning, there is a growing need for surrogate models that are both fast and precise.

Deep learning provides an alternative approach to classical fast simulations by automating the process and incorporating all of the correlations present in \geant-based simulations. In the case of calorimeter simulations, a number of machine learning approaches have been studied, including Generative Adversarial Networks (GANs)~\cite{GANs}~\cite{GanPhys2,GanPhys3,deOliveira:2017rwa,Erdmann:2018kuh,Erdmann:2018jxd,Belayneh:2019vyx,Vallecorsa:2019ked,SHiP:2019gcl,Chekalina:2018hxi,Carminati:2018khv,Vallecorsa:2018zco,Musella:2018rdi,Deja:2019vcv,ATLAS:2022jhk,ATL-SOFT-PUB-2018-001,ATLAS:2021pzo}, Variational Autoencoders~\cite{VAEs}~\cite{ATL-SOFT-PUB-2018-001,ATLAS:2022jhk,Buhmann:2021lxj,Buhmann:2021caf,Diefenbacher:2023prl}, Normalizing Flows (NFs)~\cite{NFs}~\cite{caloflow1,caloflow2,Buckley:2023rez,Krause:2022jna,Diefenbacher:2023vsw,Cresswell:2022tof, Liu:2023lnn}, and Diffusion Models~\cite{scoremodels}~\cite{mikuni:caloscore,Buhmann:2023bwk,Acosta:2023zik,Mikuni:2023tqg,Amram:2023onf}.  Other detector components have also been studied - see e.g. Ref.~\cite{Feickert:2021ajf} for an updated list.  Experiments are already starting to integrate these tools into their fast simulation workflows~\cite{ATLAS:2021pzo} and the community continues to improve upon these methods for better precision and faster generation time~\cite{calochallenge}.

Nearly every proposal so far for deep generative calorimeter simulations is a complete replacement for a \geant-based model.  This means that they must map random numbers from a known probability density (e.g. a Gaussian density) to calorimeter showers.  The challenge is that calorimeter showers do not resemble the starting distribution of the random numbers and so the neural networks have to model a complex map.   Furthermore, while deep generative models may be the only option for fast and precise simulations in small, new, or planned experiments, well-established experiments may already have an excellent classical surrogate model.  In this paper, we propose Schr\"{o}dinger bridge Quality Improvement via Refinement of Existing Lightweight Simulations (SQuIRELS), a method that uses a deep generative model for calorimeter simulation, but instead of starting from Gaussian random numbers, it starts from a classical fast simulation.

While most proposals for deep learning simulators start from a Gaussian probability density, some approaches have considered refinement methods.  One such strategy\footnote{In the special case where there is a match between the full and fast simulation events, one can also use a regression approach like the one proposed in  Ref.~\cite{Banerjee:2022gkg}.} is to reweight physics-based or deep learning-based surrogate models using neural networks~\cite{2009.03796,Nachman:2023clf,Darulis:2022brn,Adelmann:2022ozp,Winterhalder:2021ave,Das:2023ktd}.  The challenge with reweighting methods is that they dilute the statistical power of the simulated sample and fail when there are regions of non-overlapping support between the fast and full simulations.  An alternative approach is to use generative models to map one simulation into another.  Most generative models cannot naturally accommodate this setup because they require knowing the probability density of the input noise.  This is not the case for GANs, which have been studied as a refining network~\cite{Erdmann:2018kuh}. However, despite their flexibility, GANs are difficult to train due to their minimax loss function.  VAEs can be modified to use an implicit noise probability density (called the latent space)~\cite{Howard:2021pos}, but the map between samples is not constrained to be minimal and they have not been studied for refining simulations.  Models that explicitly learn the probability density or at least the gradient of the density (called the score) seem to perform the best for calorimeter simulations~\cite{calochallenge}. It is possible to use e.g. NFs to map one dataset into another, but one has to go through a known density first~\cite{klein2022flows,Mastandrea:2022vas} and the mapping is also not constrained to be minimal.  Instead, we study a modified diffusion model setup (Schr\"{o}dinger Bridge) starting from sampples that need not have an explicitly known probability density.  

This paper is organized as follows.  Section~\ref{sec:dataset} introduces the datasets we use to demonstrate SQuIRELS.  Since the fast simulations of the LHC experiments are not public or easily usable by external researchers, we create a public benchmark dataset with fast and full simulations based on \gf~\cite{Grindhammer:1989zg,Grindhammer:1993kw} and \geant, respectively.  The Schr\"{o}dinger Bridge setup is described in Sec.~\ref{sec:method}.  Results are presented in Sec.~\ref{sec:results} and the paper ends in Sec.~\ref{sec:conclusions} with conclusions and outlook.

\section{Diffusion Schr\"{o}dinger Bridges}
\label{sec:bridge}
The goal of the refinement model is to take observations $x_\alpha \sim p_\alpha$ and determine a transport function that transforms these initial set of observations into a second distribution $x_\beta \sim p_\beta$, such that samples from a fast but less accurate simulation are corrected towards a high-fidelity simulation. While the boundary conditions of this problem are fixed, the exact form of the transport function is not unique and different methods in the context of optimal transport problems~\cite{monge1781memoire} have been studied. A particular class of regularized, optimal transport problems, showing promissing results for high-dimensional data, are solutions of the Schr\"{o}dinger Bridge (SB)~\cite{schrodinger1931umkehrung} problem. Given an initial  Markov chain with initial probability density $p_0(x_0)=p_\alpha$, we can sequentially apply perturbations with transition kernels $p_{k+1|k}$ such that after $N$ perturbations we arrive at the following joint probability density function:
\begin{equation}
     p(x_0,N) = p_0(x_0)\,\prod_{k=0}^{N-1}p_{k+1|k}(x_{k+1}|x_k).
     \label{eq:ref}
\end{equation}

While the starting distribution satisfies the initial condition of the data we want to correct, the end result of this diffusion process does not necessarily match the requirement $p_N(x_N)=p_\beta$. Given this reference path of probability densities, however, the SB problem is to identify what path $\pi^*$  satifies the boundary conditions while also being closest to the reference path. The path $\pi^*$ then satisfies:
\begin{equation}
    \pi^* = \mathrm{arg min}\left\{ \mathrm{KL}(\pi|p): \pi_0 = p_\alpha, \pi_N = p_\beta\right\},
\end{equation}
where the Kullback–Leibler divergence is the metric used to determine the distance between possible paths. A numerical solution for this problem was proposed in the Iterative Proportional Fitting (IPF) algorithm~\cite{fortet1940resolution,kullback1968probability,chen2020optimal}. IPF uses a recursion rule to solve half-bridges by starting from the reference dynamics in Eq.~\ref{eq:ref}, $\pi^0 = p$, and following:
\begin{align}
    \label{eq:IPF}
    \pi^{2n+1} &= \text{arg min}\left\{ \text{KL}(\pi|\pi^{2n}):  \pi_N = p_\beta\right\}\\    
    \pi^{2n+2} &= \text{arg min}\left\{ \text{KL}(\pi|\pi^{2n+1}):  \pi_0 = p_\alpha\right\} \nonumber.        
\end{align}
The standard methodology to iterate the paths $\pi$ in Eq.~\ref{eq:IPF} is to obtain representations of $\pi^n$ by updating the joint density density $p$ with potential functions~\cite{bernton2019schr,chen2016entropic,pavon2018data}, often hard to approximate in the context of generative models where only samples from unknown densities are available. The authors of Ref.~\cite{de2021diffusion} propose to instead approximate the IPF algorithm using a different methodology that does not require explicit knowledge of the density functions and can be used to directly transport samples from one density to the other. Given forward and backwards Gaussian transition kernels $p^n_{k+1|k}(x_{k+1}|x_k) = \mathcal{N}(x_{k+1};F^n_k(x_k),2\gamma_{k+1})$ and $q^n_{k|k+1}(x_{k}|x_{k+1}) = \mathcal{N}(x_{k};B^n_{k+1}(x_{k+1}),2\gamma_{k+1})$ with perturbation size $\gamma_k$, we have a different recursive rule:
\begin{align}
    B^n_{k+1}&=
    \text{argmin}_B\mathbb{E}\left\| B(x_{k+1}) - x_{k+1} - F_k^n(x_k) + F^n_k(x_{k+1})\right\|^2 \label{eq:bn}\\
    F^{n+1}_{k}&=\text{argmin}_F\mathbb{E}\left\| F(x_{k}) - x_{k} - B_{k+1}^n(x_{k+1}) +B^n_{k+1}(x_{k})\right\|^2. \label{eq:fn}
\end{align}
In practice, neural networks are trained to approximate the forward and backward drift functions $F^n_k(x)$ and $B^n_k(x)$, with trainable parameters $\sigma^n$ and $\theta^n$, respectively. Starting from samples $x_0\sim p_\alpha$ we define 
\begin{equation}
    x_{k+1} = F_\sigma^n(k,x_{k}) + \sqrt{2\gamma_{k+1}}Z,
    \label{eq:for}
\end{equation}
where $Z\sim \mathcal{N}(0,1)$. From this forward process, we define the loss function $l_n^b$ by minimizing Eq.~\ref{eq:bn}, where the forward network $F_\sigma^n$ is fixed and only the backward network $B_\theta^n$ is updated during backpropagation. After convergence, we can use the trained $B_\theta^n$ to sample from the reverse process such that:
\begin{equation}
    x_{k-1} = B_\theta^n(k,x_{k}) + \sqrt{2\gamma_{k}}\tilde{Z},
    \label{eq:back}
\end{equation}
where $x_N\sim p_\beta$ and $\tilde{Z}\sim \mathcal{N}(0,1)$. These new trajectories are then used to calculate the loss function $l_n^f$ based on Eq.~\ref{eq:fn}, where $B_\theta^n$ is now fixed and only $F_\sigma^n$ is updated. After each iteration $n$, a new pair of neural networks $B_\theta^n$ and $F_\sigma^n$ are trained. However, since each iteration only represents a small correction from the previous pair of trained networks, these can be initialized based on the previous trained models for faster convergence. After training, samples can be transported back and forth between distributions using Eqs.~\ref{eq:for} and \ref{eq:back}.

\section{Data Sets}
\label{sec:dataset}

In order to train and benchmark SQuIRELS, we generate two groups of data sets of electron showers in an electromagnetic calorimeter. The first group is simulated using the parameterized fast simulation setup \gf~\cite{Grindhammer:1989zg, Grindhammer:1993kw}. This group presents the showers that we intend to refine. The second group presents the targets for this refinement and is simulated using \geant~\cite{geant4}, version 11-01-patch-02, with the FTFP\_BERT physics list.

The calorimeter geometry and setup used for both full- and fast simulation is based on the \gf example provided by \geant. The calorimeter consists of 100 crystals of Lead-Tungstate (PbWO4) arranged in a regular $10\times10$ grid pattern, with each crystal having a height and width of 3~cm and a length of 24~cm. 

We use the default \gf parametrization functions, and no finetuning or correction is performed to improve the performance. The parametrization stepsize is unmodified from its default value of 0.1 radiation lengths. For these reasons, it is important to note that this work should not be seen as representative of what performance and accuracy can be achieved when using parameterized fast simulation. Our goal is not to investigate the \gf itself but to demonstrate what kinds of mis-modeling SQuIRELS is capable of correcting. 

The training set consists of 100k \gf electron showers and 100k \geant electron showers, simulated with incident particle energies randomly sampled between 10 and 100 GeV. To allow for a one-to-one mapping between fast and full simulation, the sampled incident energies are ensured to be identical between the \gf and \geant runs. In addition to the training set, four further sets, each containing 100k fast- and full-sim showers, are generated, one set with energies between 10 and 100 GeV, and three sets with fixed incident energies of 20~GeV, 50~GeV, and 80~GeV. These additional datasets are used for evaluation.

\section{Schr\"{o}dinger Bridge models}
\label{sec:method}

The goal of SQuIRELS is to refine a set of starting samples, made up of \gf showers and called $X_{GF}$, to match a  set of target samples, consisting of \geant showers and called $X_{G4}$. Individual samples from the sets are denoted as $x_{GF} \in X_{GF}$ and $x_{G4} \in X_{G4}$ with both $x_{G4} \in \mathbb{R}^{10\times10}$ and  $x_{GF} \in \mathbb{R}^{10\times10}$. 

To achieve the desired mapping, SQuIRELS takes notes from previous ML calorimeter modeling approaches~\cite{caloflow1,caloflow2} and splits the refinement into a two steps process. The first step consists of a one-dimensional Schr\"{o}dinger Bridge, referred to as SQuIRELS-energy (SQuIRELS-E), that maps the total energy sum $e_{GF} = \sum_{i=1,j=1}^{10} x_{i,j,GF}$  of a given \gf shower to that of an equivalent \geant shower with $e_{G4} = \sum_{i=1,j=1}^{10} x_{i,j,G4}$. Great care is taken to ensure both $e_{GF}$ and $e_{G4}$ were simulated using the same incident particle energy. SQuIRELS-E is conditioned on this energy of the incident electron that causes the shower, in addition to the standard conditioning on the timesteps. The model at the core of SqUIRELS-E is a fully connected network consisting of 3 encoding networks with 3 layers of 256, 256, and 128 nodes. The inputs to the encoding networks are the current datapoint, the current timestep, and the incident particle energy, respectively. The timestep is processed using sinusoidal embeddings~\cite{DBLP:journals/corr/VaswaniSPUJGKP17}, which are common in diffusion models. The outputs of the 3 encoders are concatenated along with an additional instance of the current datapoint and passed into a decoding network with 3 layers of 256, 256, and 1 nodes. LeakyReLU activations are used throughout the SQuIRELS-E network, except for the final output layers of the subnetworks. 

The inputs to SQuIRELS E are preprocessed by first dividing the $e_{GF}$ and $e_{G4}$ by the energy of their associated incident particles, and then modifying the distributions to have a mean of 0 and a width of 1. This is achieved by shifting and rescaling the distributions by two sets of parameters, $\mu_{GF}$ and $\sigma_{GF}$, for \gf energy sums, and $\mu_{G4}$ and $\sigma_{G4}$ for \geant energy sums, respectively. It should be noted here, that when undoing this preprocessing, the set of parameters associated with the target distribution is used, so that a given \gf energy $e_{GF}$ is first shifted and scaled by $\mu_{GF}$ and $\sigma_{GF}$, then refined by SQuIRELS-E, and then inversely scaled by $\mu_{G4}$ and $\sigma_{G4}$.

In the second step, a larger Schr\"{o}dinger Bridge, simply called SQuIRELS, is used to refine the full $10\times10$ sized calorimeter showers. In addition to being conditioned on the sinusoidal embedded timestep, the SQuIRELS model is further conditioned on the energy of the incident particle, the energy sum $e_{GF}$ of the current shower, and the energy sum that the refined shower is supposed to have, $e_{G4}$. During the model training, $e_{GF}$ and $e_{G4}$ are taken from the training data, while during generation, $e_{G4}$ is provided by SQuIRELS-E. SQuIRELS leverages the lack of network architecture restrictions of the Schr\"{o}dinger Bridge framework and makes use of a convolutional neural network. Similarly to the SQuIRELS-E model, the SQuIRELS model comprises an encoding network and a decoding network. The encoder consists of 4 convolutional layers with 32 output filters, $3\times3$-sized kernels, and `same'-padding each. The current data point is passed to the encoder, the output of which is subsequently concatenated with the conditional inputs, which are broadcasted to a $10\times10$ shape, and an additional instance of the current data point. This then comprises the input to the decoder network, which consists of another 4 convolutional layers with 32, 32, 32, and 1 output filters, $3\times3$-sized kernels, and `same'-padding each, thereby mirroring the encoder. As was done in SQuIRELS-E, LeakyReLU activations are used throughout the SQuIRELS network, except in the final output layers of the subnetworks, which have no activation. 

Inputs to SQuIRELS are pre-processed similarly to the SQuIRELS-E inputs. First, all pixel values are divided by the total energy in the shower, normalizing the sum of all pixels to 1. Then, each of the 100-pixel positions is individually shifted and scaled such that the means and standard deviations across all showers in the training set for a given pixel position are 0 and 1, respectively. As with SQuIRELS-E, the parameters of the target distribution are used to reverse this pre-processing after SQuIRELS performs its refinement. The energy sum needed to undo the normalization is again provided by SQuIRELS-E.

All models were implemented using \textsc{PyTorch}~\cite{pytorch} version 2.0.1. and were trained using the Adam~\cite{adam} optimizer with a learning rate of $10^{-5}$ and default momentum parameters. Both models employ a flat noise schedule with $20$ steps of $0.001$ and make use of exponential moving averaging (EMA) with a decay factor of $0.95$. The EMA code was adapted from Ref.~\cite{EMAcode}. The implementation for both SQuIRELS bridges is adapted from Ref.~\cite{de2021diffusion}. 

\section{Results}
\label{sec:results}

\subsection{SQuIRELS-E}

We first examine the performance of the SqUIRELS energy model. This allows us to test if the refinement step performs sufficiently well and serves as a small-scale demonstration of the Schr\"{o}dinger Bridge.

For all examined observables, we also calculate the Earth mover's distance (EMD) directly on the one-dimensional distributions. The EMD calculation uses a bootstrapping approach~\cite{10.1214/aos/1176344552} to estimate the uncertainty of the EMD value. This does, however, mean that even for two identical data sets, the calculated EMD is not precisely 0.0, as the bootstrapped instances of the sets will still differ. To account for this lower limit on the EMD value, we include the EMD calculated between the references \geant sample and itself. These quantitative results are shown in the legends of the figures and, for the full SQuIRELS setup, are summarized in Table~\ref{tab:EMD}.

\begin{figure*}[ht]
\centering
    \includegraphics[width=0.3\textwidth]{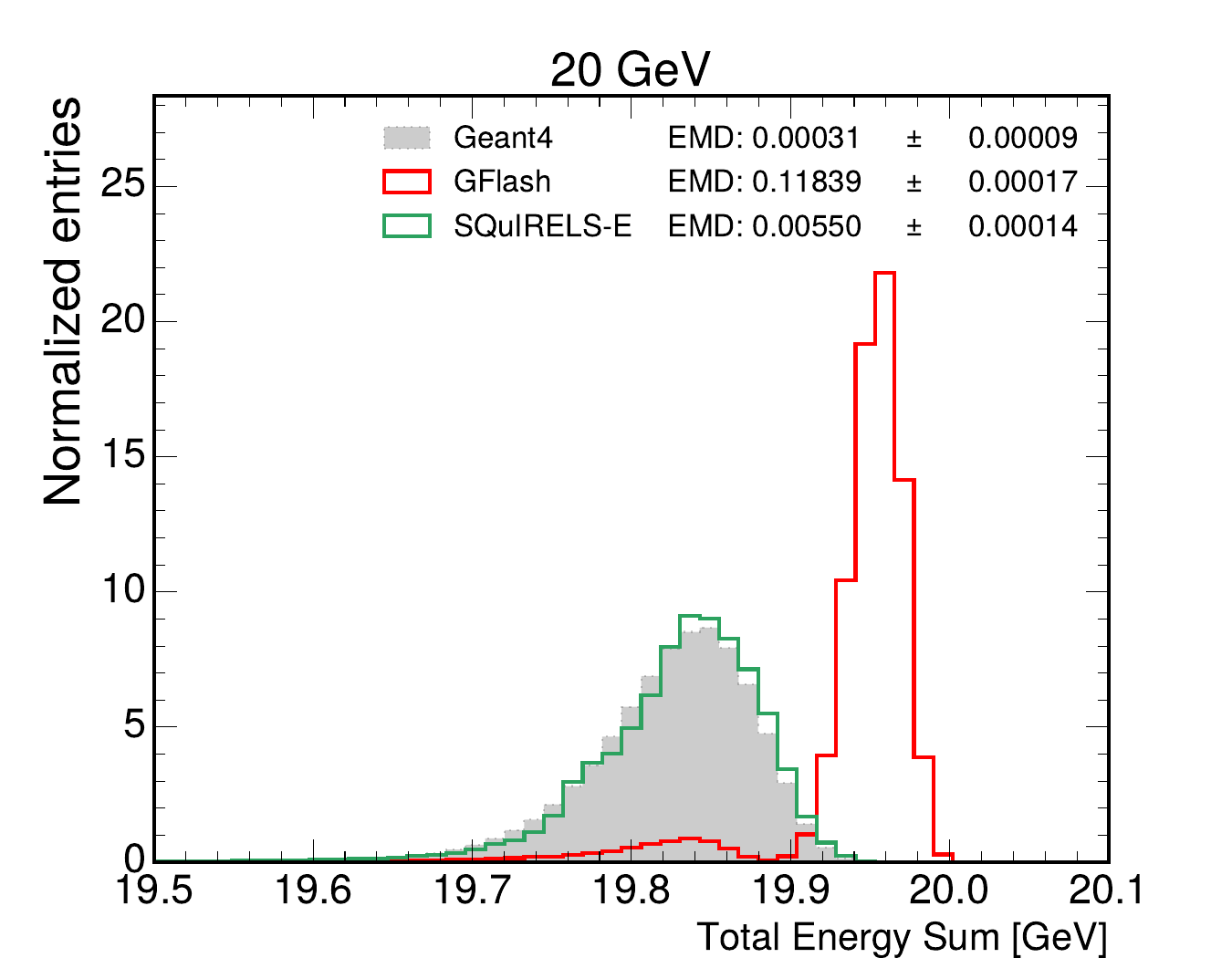}
    \includegraphics[width=0.3\textwidth]{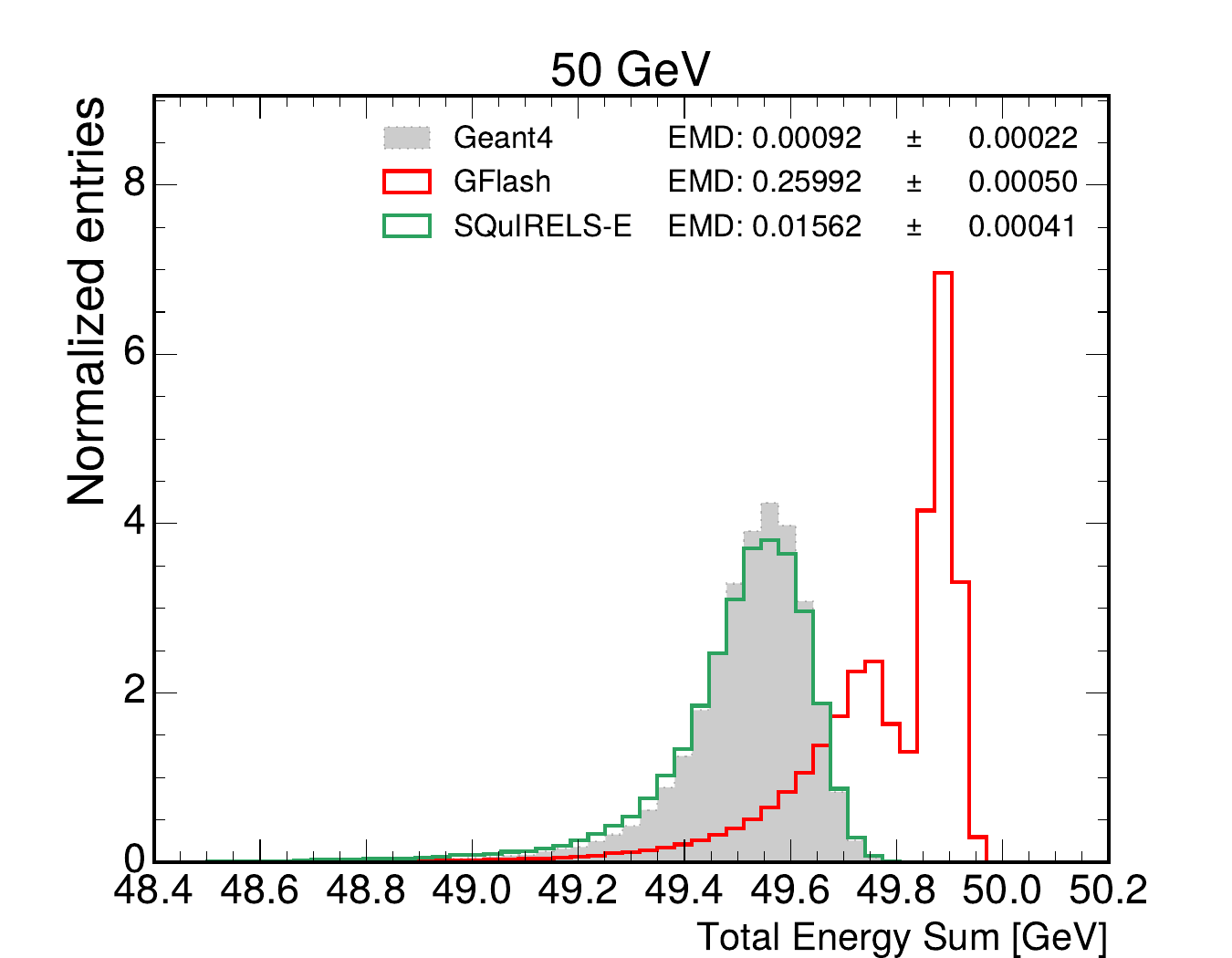}
    \includegraphics[width=0.3\textwidth]{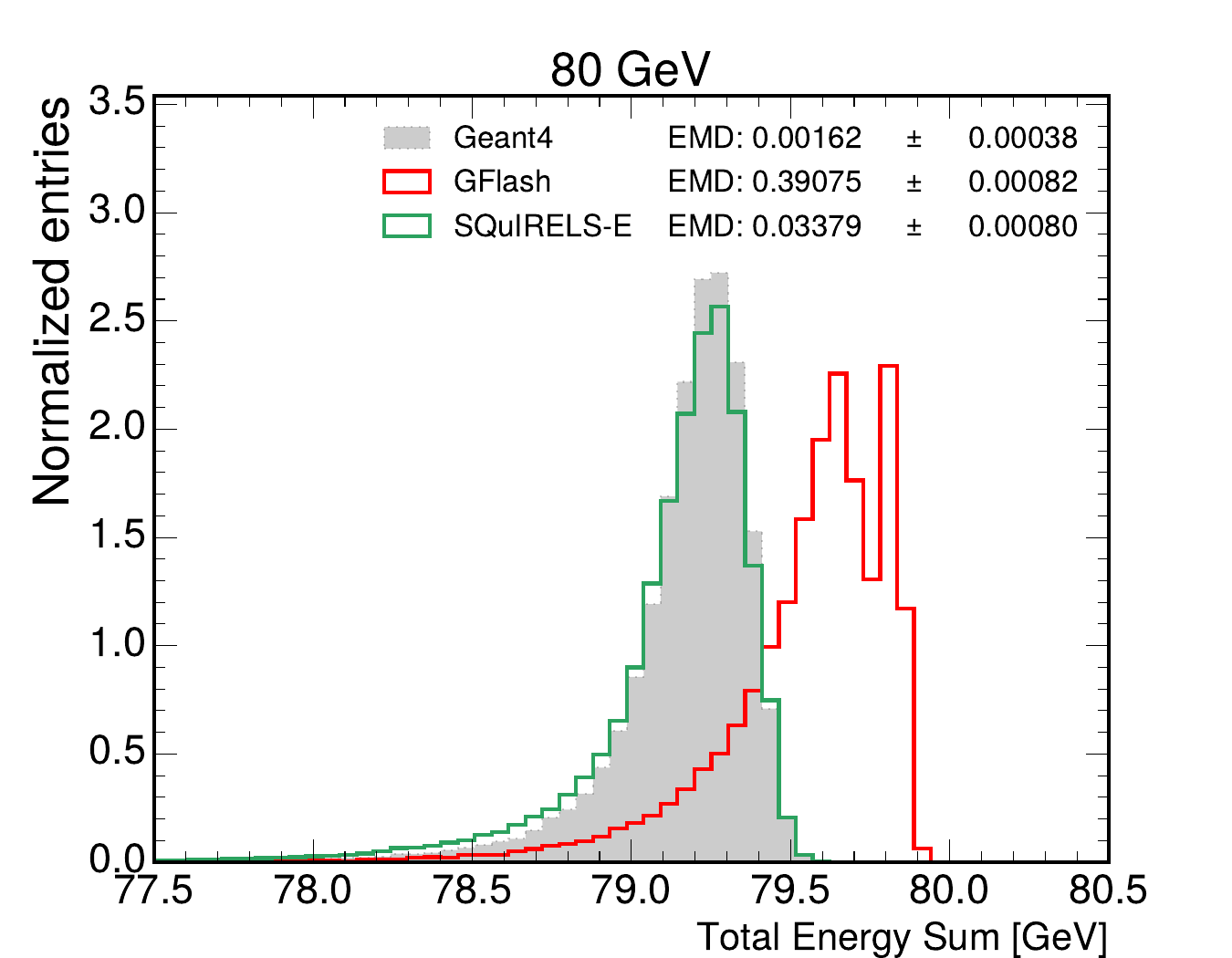}    
\caption{Comparison between full simulation (labeled \geant), fast simulation (labeled \gf), and the one-dimensional energy sum Schr\"{o}dinger Bridge-refined fast simulation (labeled SQuIRELS-E). The panels show (from left to right) the total deposited energy for 20 GeV, 50 GeV, and 80 GeV electrons respectively. EMD values provide a quantitative agreement score between the reference \geant and the 3 methods, see text for more detail.}
\label{fig:EB_energy_sum}
\end{figure*}

\begin{figure*}[ht]
\centering
    \includegraphics[width=0.3\textwidth]{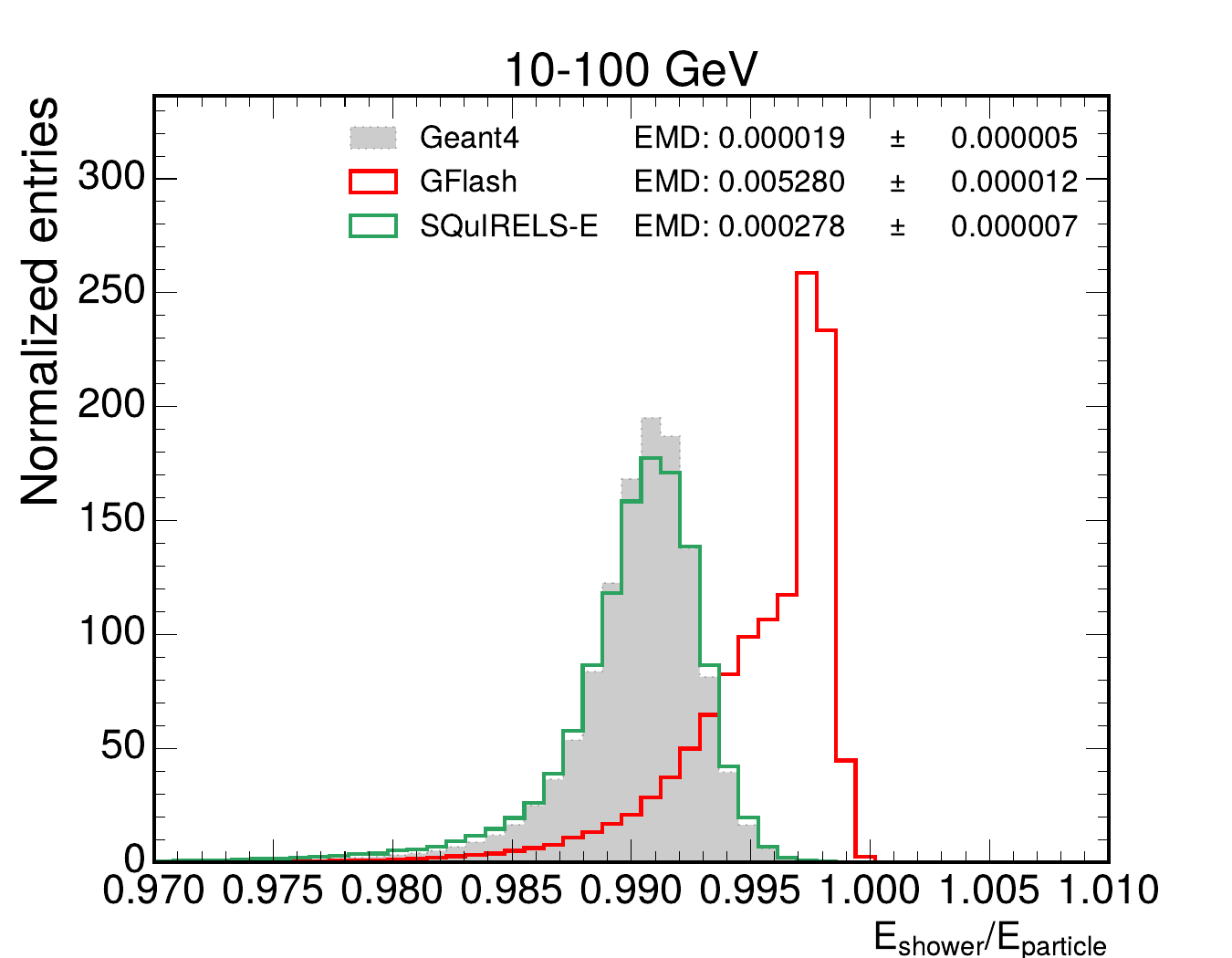}
    \includegraphics[width=0.3\textwidth]{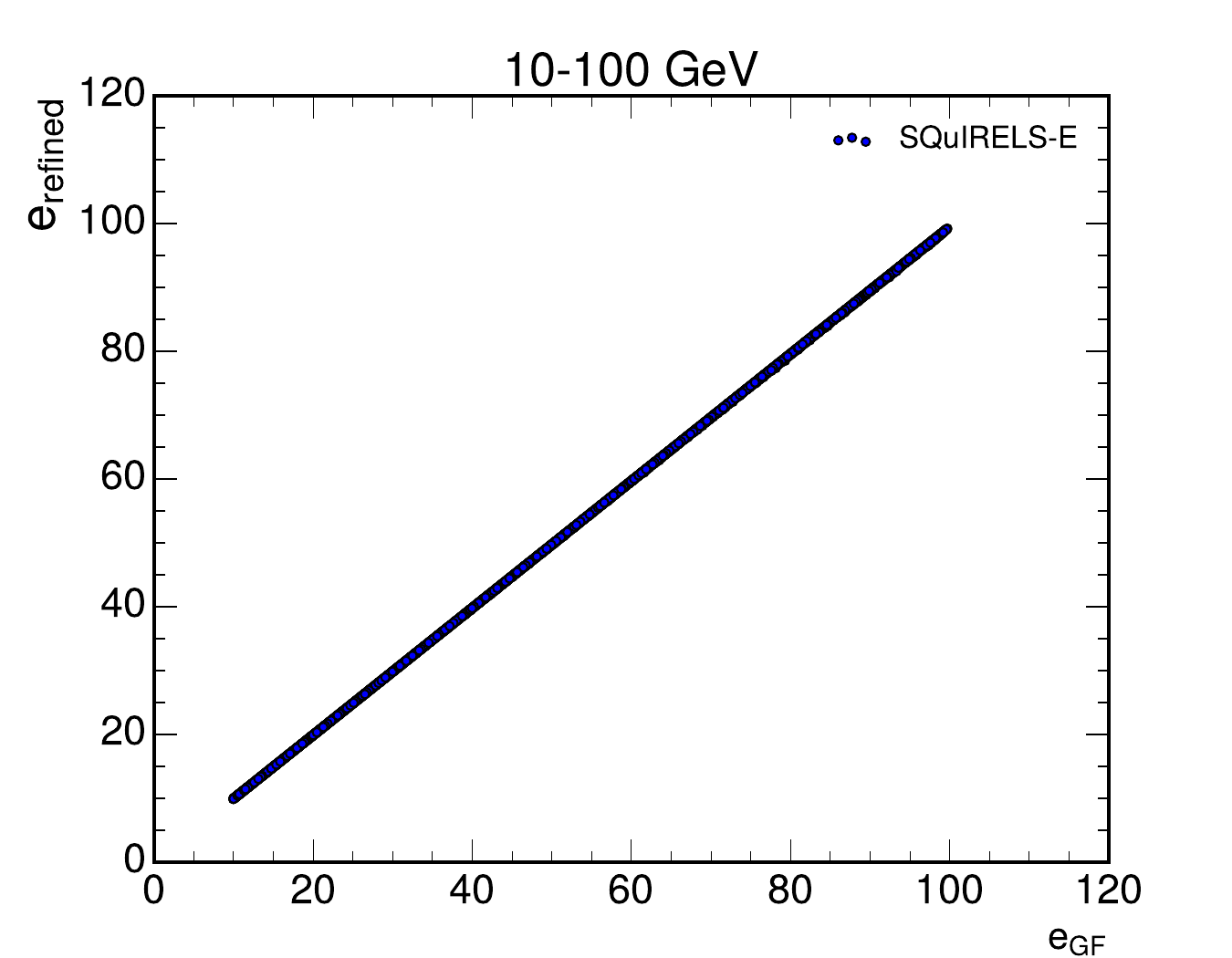}
    \includegraphics[width=0.3\textwidth]{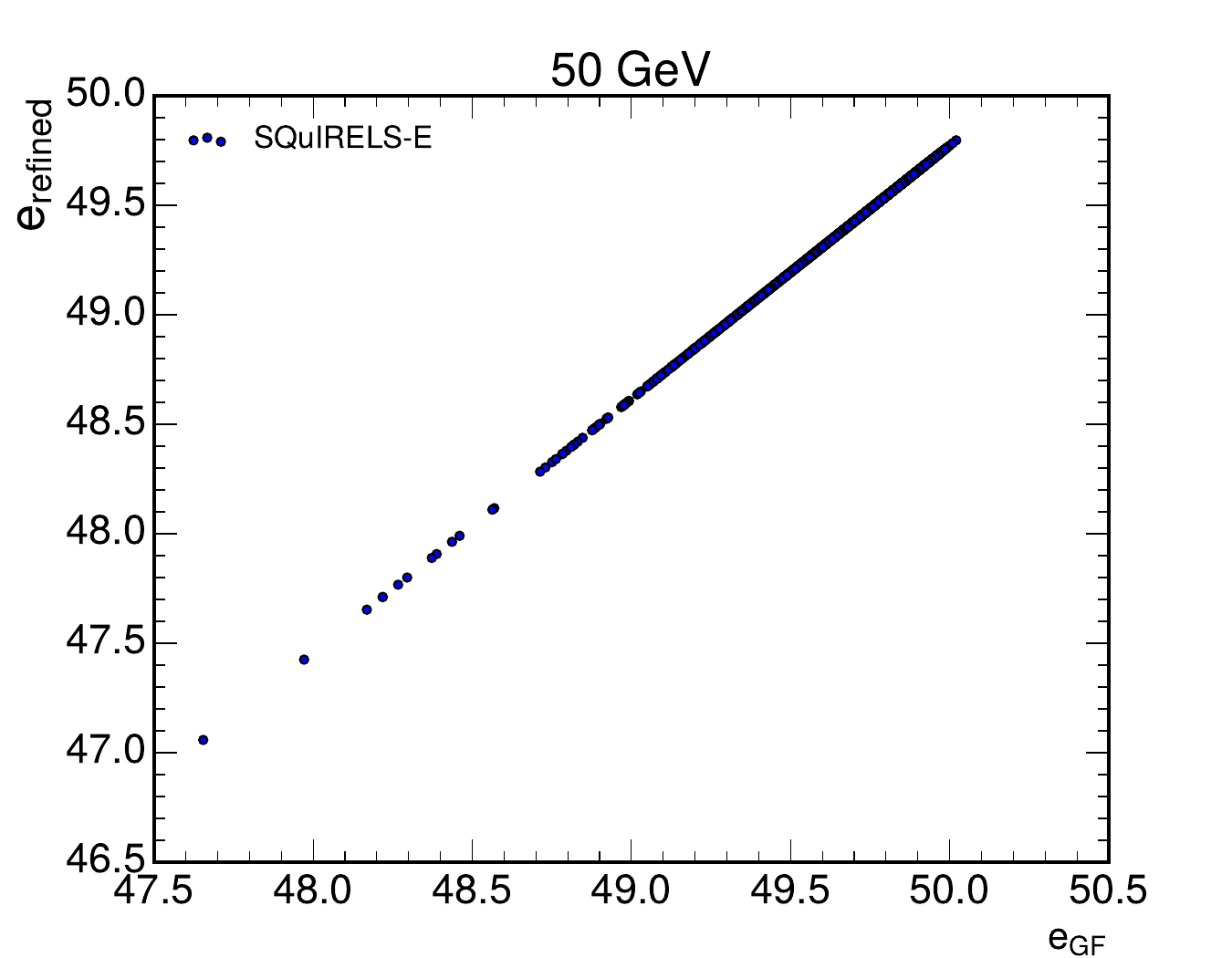}  
\caption{Comparison between full simulation (labeled \geant), fast simulation (labeled \gf), and the one-dimensional energy sum Schr\"{o}dinger Bridge-refined fast simulation (labeled SQuIRELS-E). The leftmost panel shows the ratio between the total energy deposited within the calorimeter and the energy of the incident electron. The remaining two panels show scatter plots between the inputs and outputs of SQuIRELS-E, for 10-100~GeV, and for 50~GeV showers. EMD values provide a quantitative agreement score between the reference \geant and the 3 methods, see text for more detail.}
\label{fig:EB_energy_sum_Conditioning}
\end{figure*}

Fig.~\ref{fig:EB_energy_sum} shows the energy sums over all pixels for \geant ($e_{G4}$, grey), \gf ($e_{GF}$, red), and the refined energy sums produced by the SqUIRELS-E (green). The three different panels correspond to the three single incident particle energy test sets, with the lefthand panel, showing energy sums for 20~GeV electron showers, the center panel showing 50~GeV showers, and the righthand panel showing 80~GeV ones. The systematic mismatch between the \geant and \gf can be seen for all three particle energies, with \gf showers having on average more deposited energy and the energy sum distribution displaying a noticeable double peak structure. While we do not aim to elucidate the origin of these mismatches, they do serve as an illustrative example of the power of Schr\"{o}dinger Bridges. The bridge, as can be seen across all the shown particle energies, achieves a very high level of agreement with the \geant ground truth, both in terms of the energy peak position and in terms of its shape. 

Further details on how the energy sums are refined by SqUIRELS-E can be seen in Fig.~\ref{fig:EB_energy_sum_Conditioning}. The leftmost panel of the figure compares the shower energy sums across the whole 10~GeV to 100~GeV range, normalized by the energy of the incident particle. We again see notable deviations between \geant and \gf, while the SqUIRELS-E output matches \geant very well. The remaining two panels show scatter plots of the SQuIRELS-E inputs (the \gf energy sums) and its outputs. The center panel shows this for showers produced by 10~GeV to 100~GeV particles, while the rightmost plot shows 50 GeV particle showers only. In both cases, we can see a nearly linear relationship between input and output energies. This indicates that model does, in fact, learn a near-optimal mapping between original energy and refined energy, where high energetic \gf showers are mapped to high energetic refined showers, and vice versa for low energetic showers.

\begin{figure*}[ht]
\centering
    \includegraphics[width=0.3\textwidth]{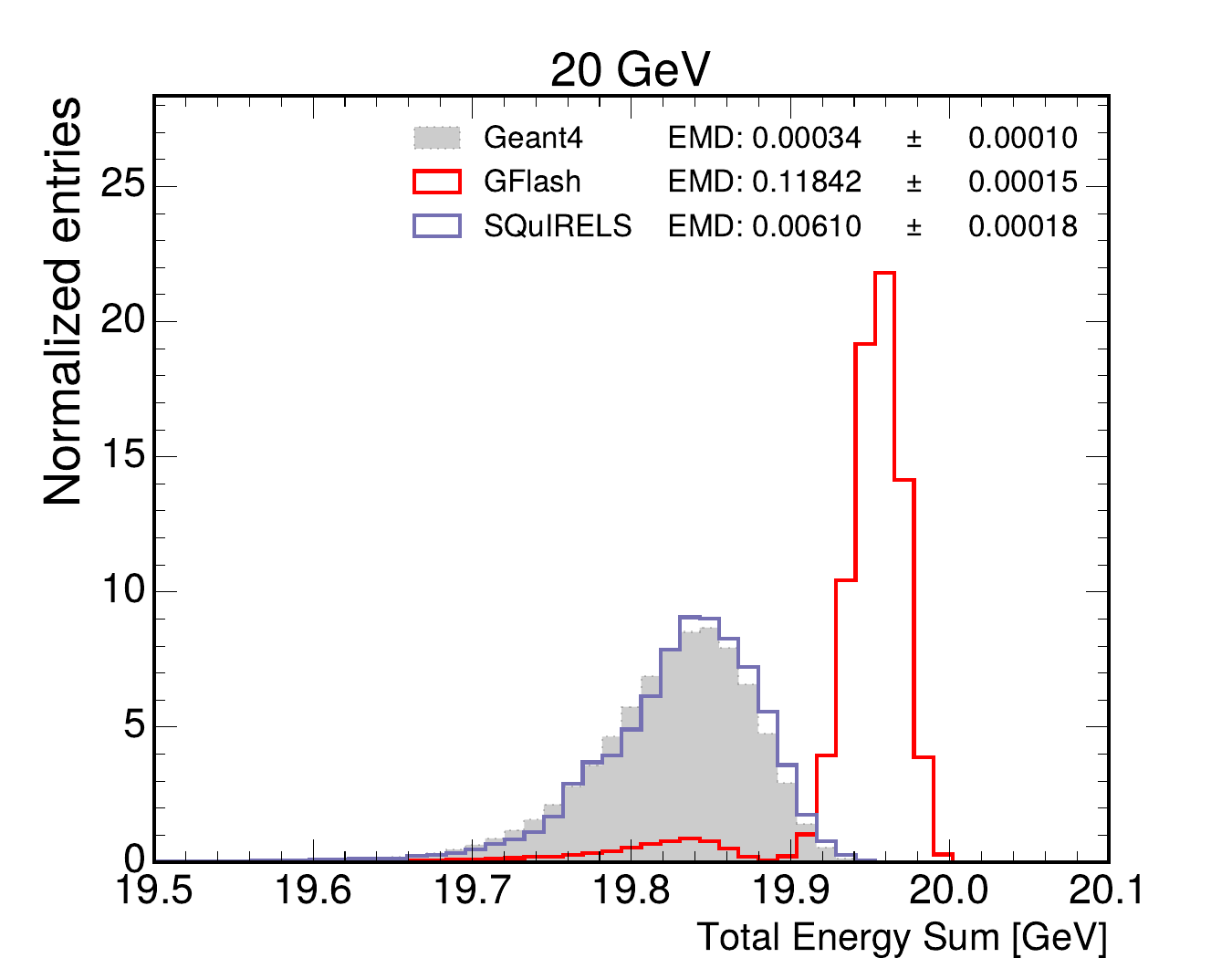}
    \includegraphics[width=0.3\textwidth]{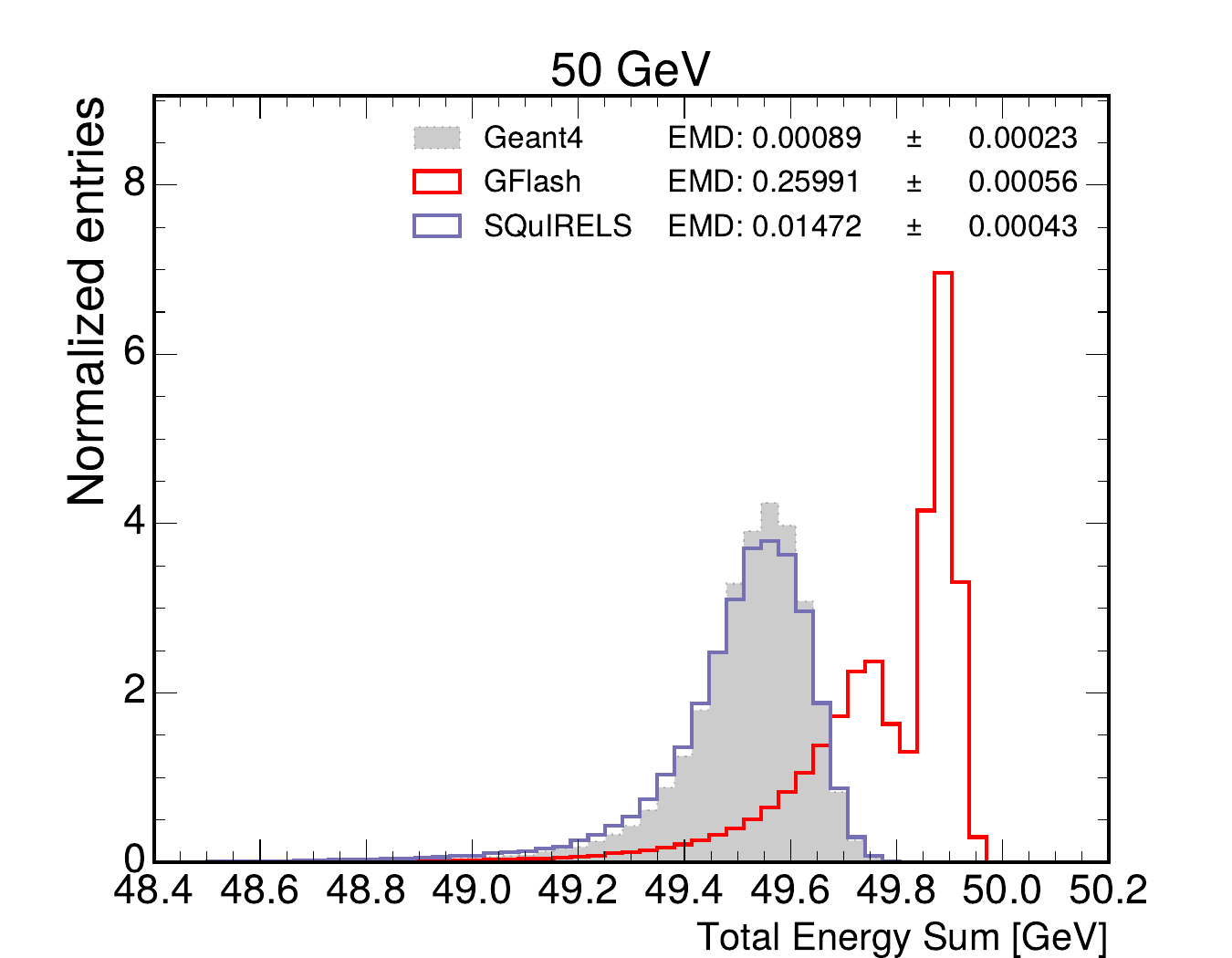}
    \includegraphics[width=0.3\textwidth]{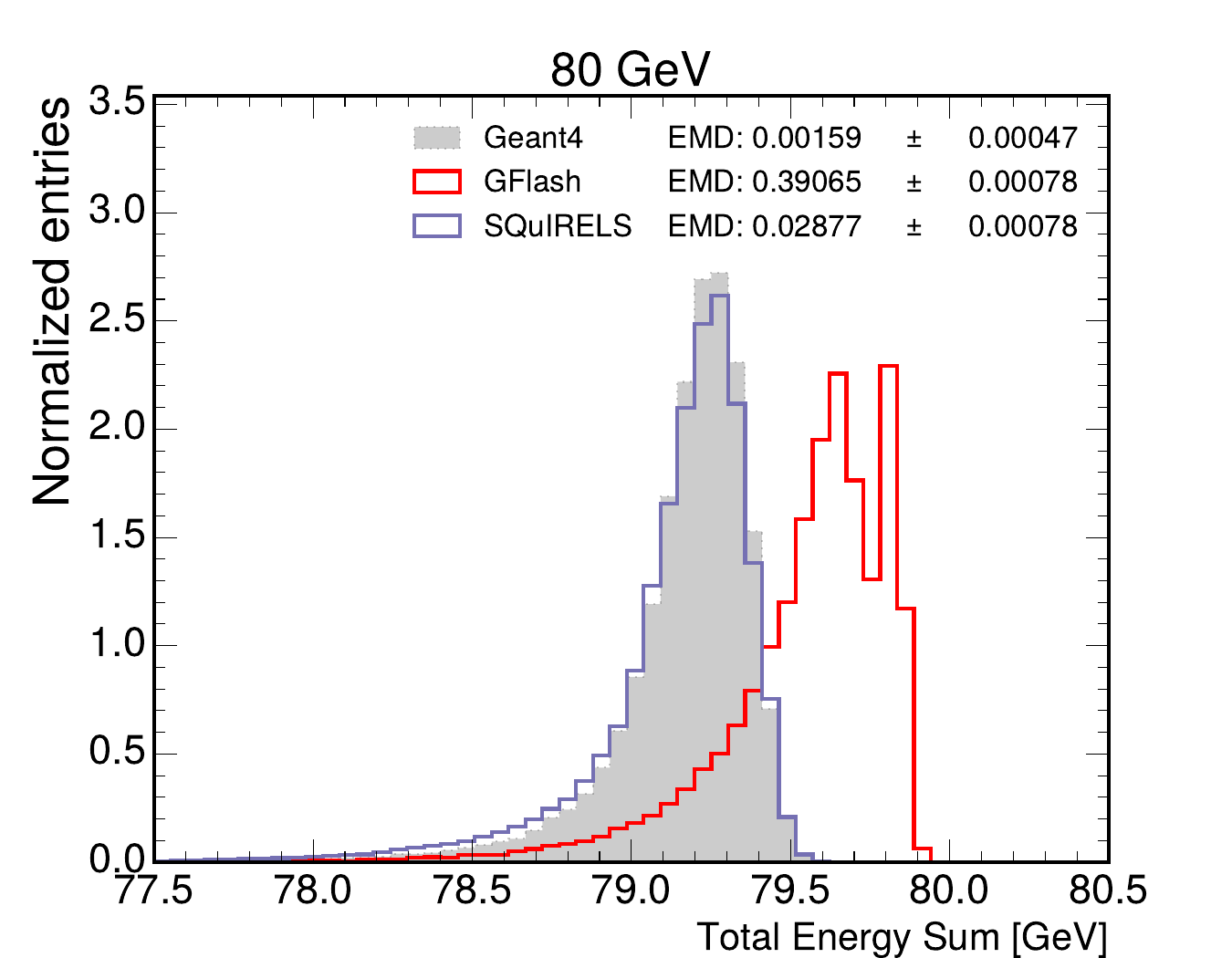}
    \includegraphics[width=0.3\textwidth]{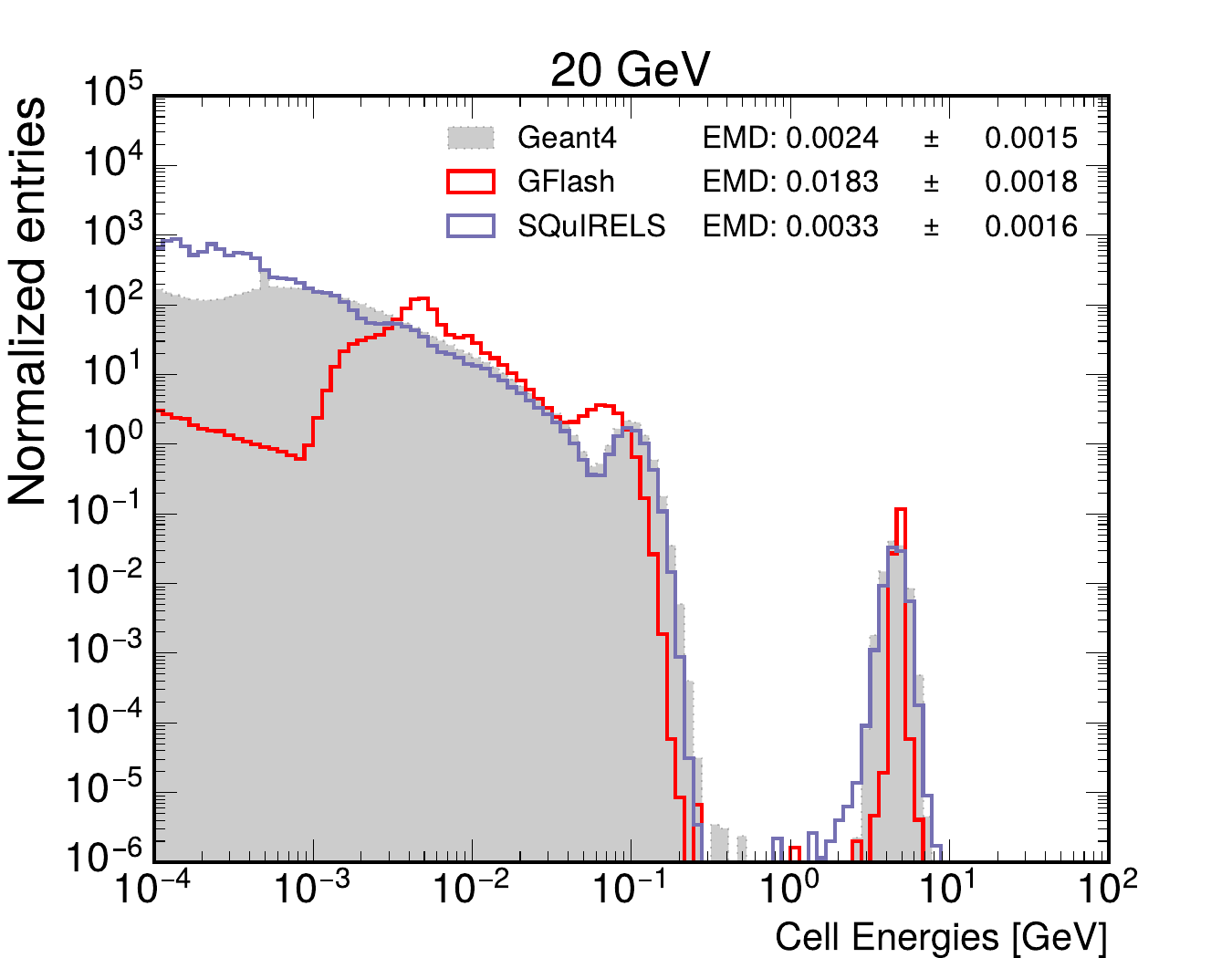}
    \includegraphics[width=0.3\textwidth]{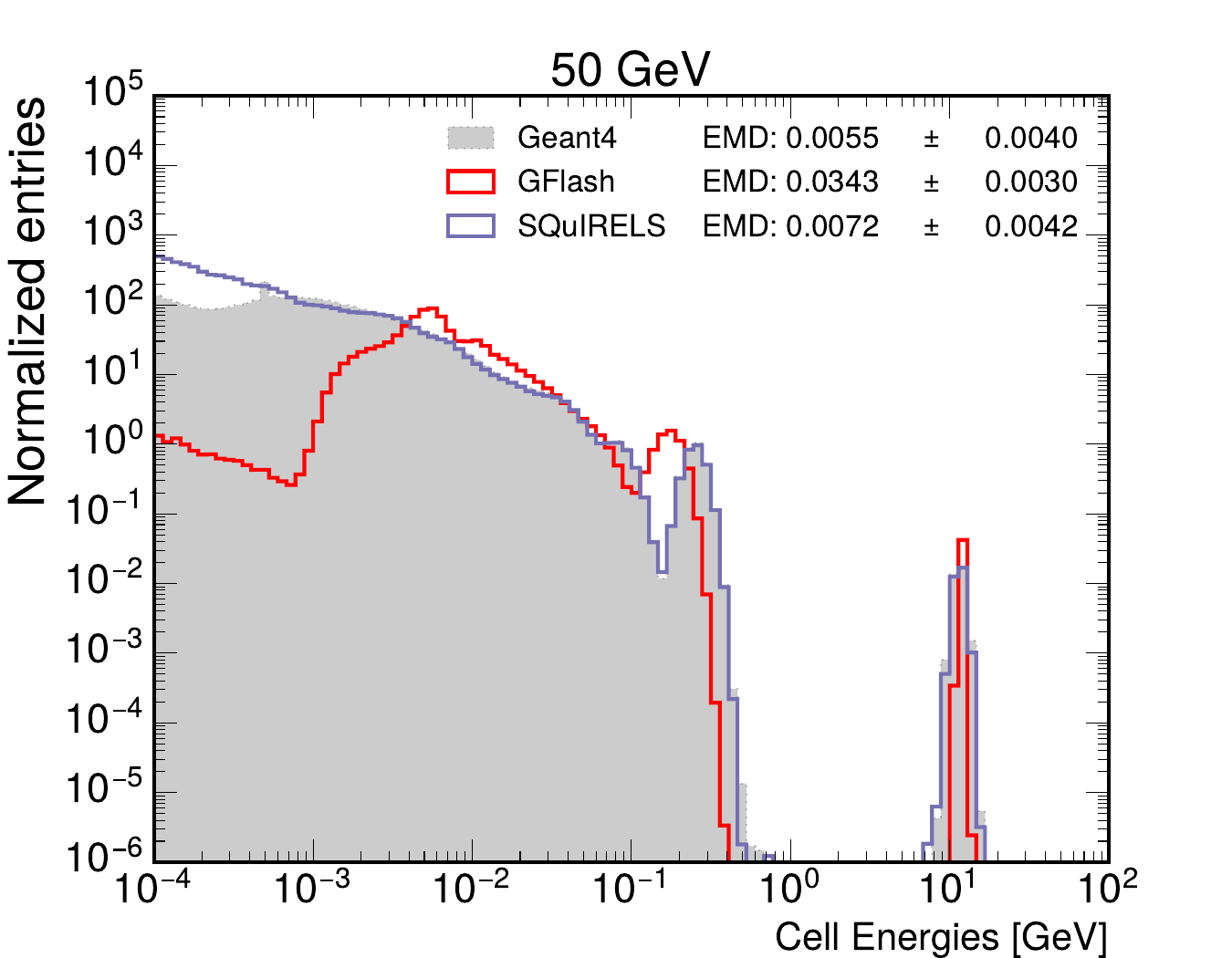}
    \includegraphics[width=0.3\textwidth]{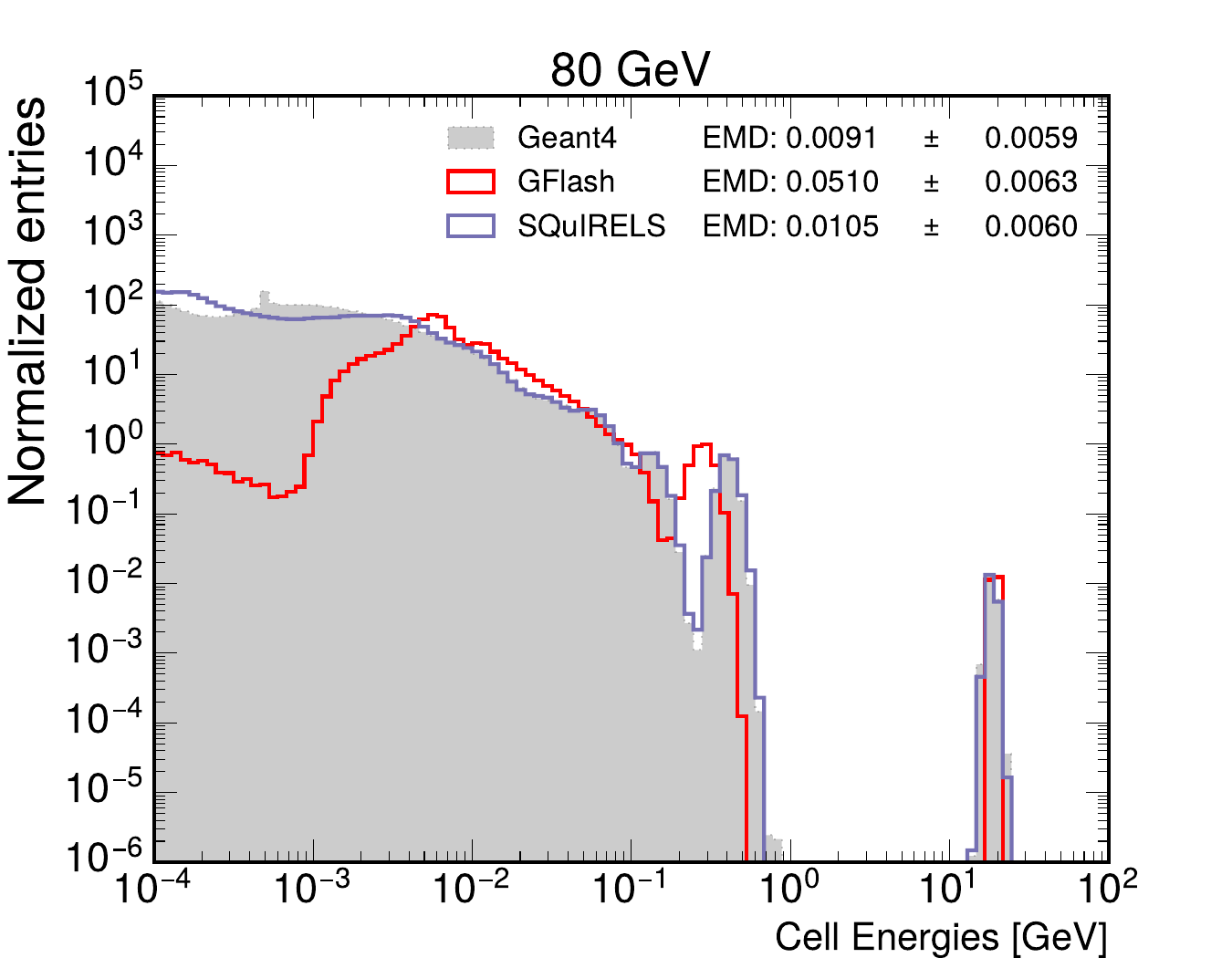}
    
\caption{Comparison between full simulation (labeled \geant), fast simulation (labeled \gf), and the Schr\"{o}dinger Bridge-refined fast shower simulation (labeled SQuIRELS). The panels show (from left to right) the total deposited energy for 20 GeV, 50 GeV, and 80 GeV electrons respectively. The top row shows the total energy sum over all pixels in the calorimeter, while the lower row shows the energy spectra of the calorimeter cells. EMD values provide a quantitative agreement score between the reference \geant and the 3 methods, see text for more detail. Note that for computational reasons, the EMD calculation of the energy spectra was limited to the plotted range and limited to 100,000 points}
\label{fig:SB_esum}
\end{figure*}

\subsection{SQuIRELS Full}

Next, we move to the full SqUIRELS Shower Bridge, which builds on the SQuIRELS-E model and uses its output as a conditional input to refine the full $10\times10$ calorimeter showers. To gauge the SQuIRELS performance, we examine a series of physics observables. As in the previous comparisons, we make use of the 20~GeV, 50~GeV, and 80~GeV single particle energy data set to serve as the basis of our comparison between \geant (grey), \gf (red), and SqUIRELS (blue). 

The top row of Fig.~\ref{fig:SB_esum} shows the sum of all deposited energy in all cells for a given set of incident particle energies. The showers produced by SQuIRELS are rescaled to match the energy sum given by SQuIRELS-E, which can be seen in the similarity between the top row plots and Fig~\ref{fig:EB_energy_sum}. As such, this presents a crosscheck that the energy rescaling works as intended and high level of agreement in the energy sum distribution is maintained even in the SQuIRELS model. 

\begin{figure*}[ht]
\centering
    \includegraphics[width=0.3\textwidth]{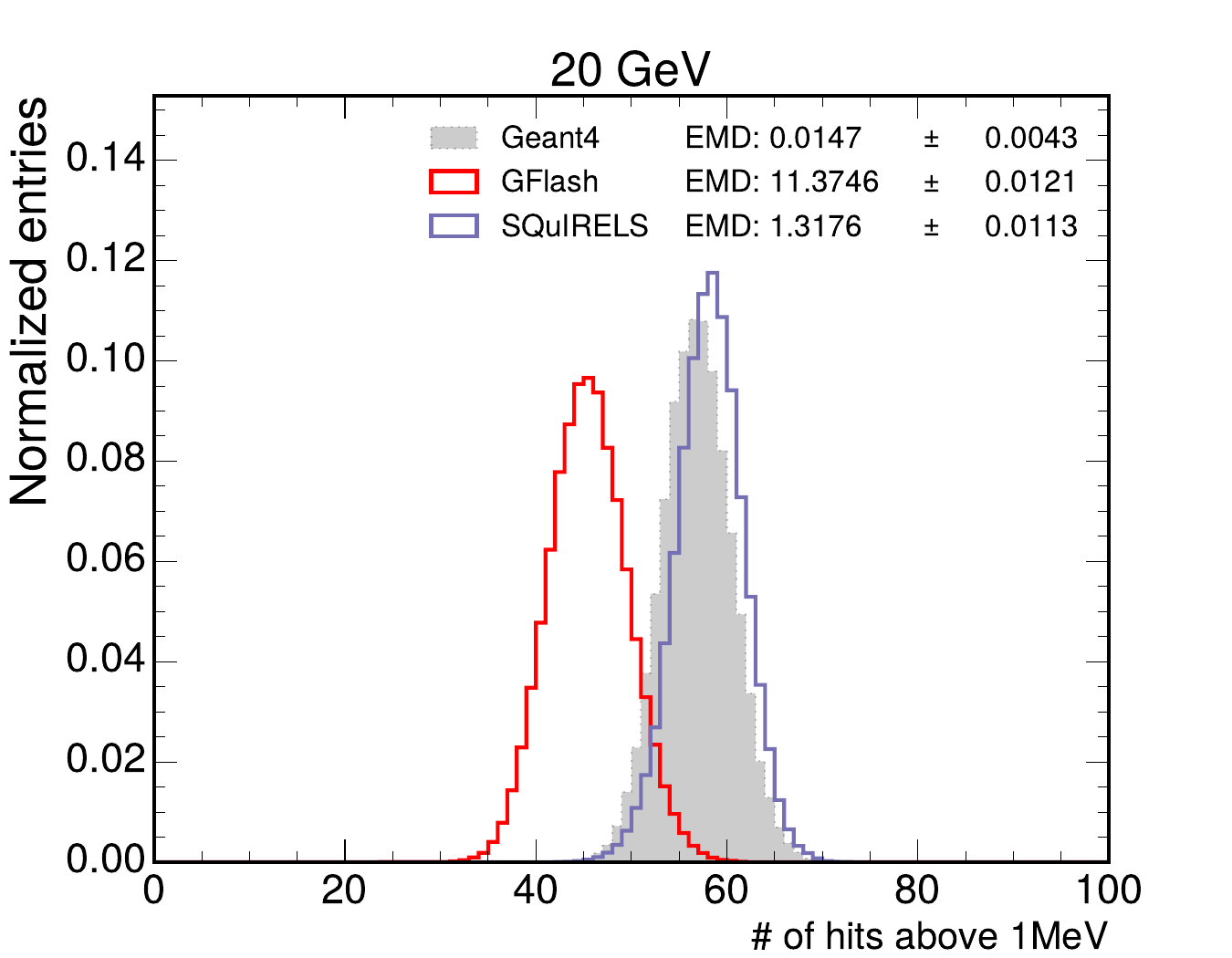}
    \includegraphics[width=0.3\textwidth]{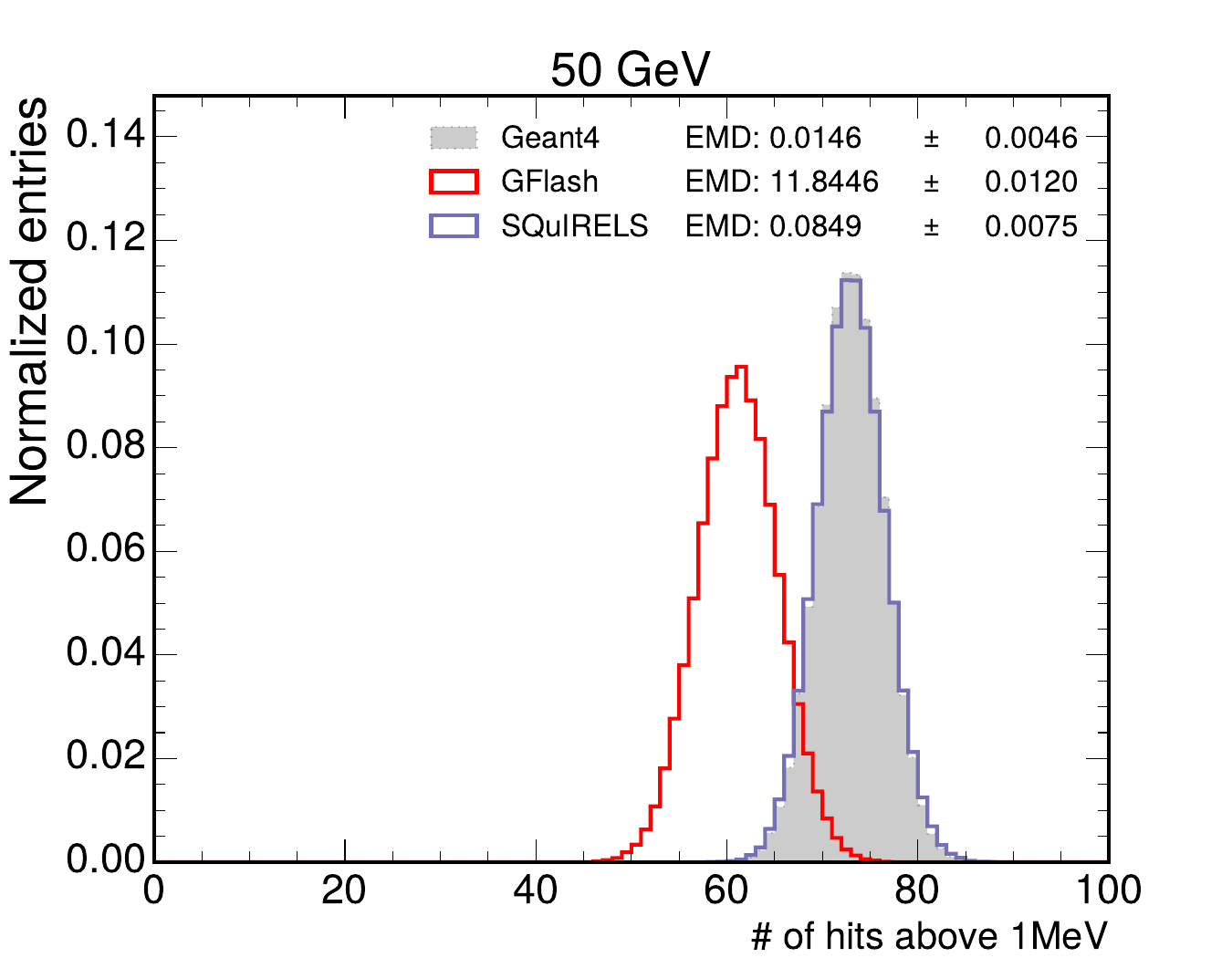}
    \includegraphics[width=0.3\textwidth]{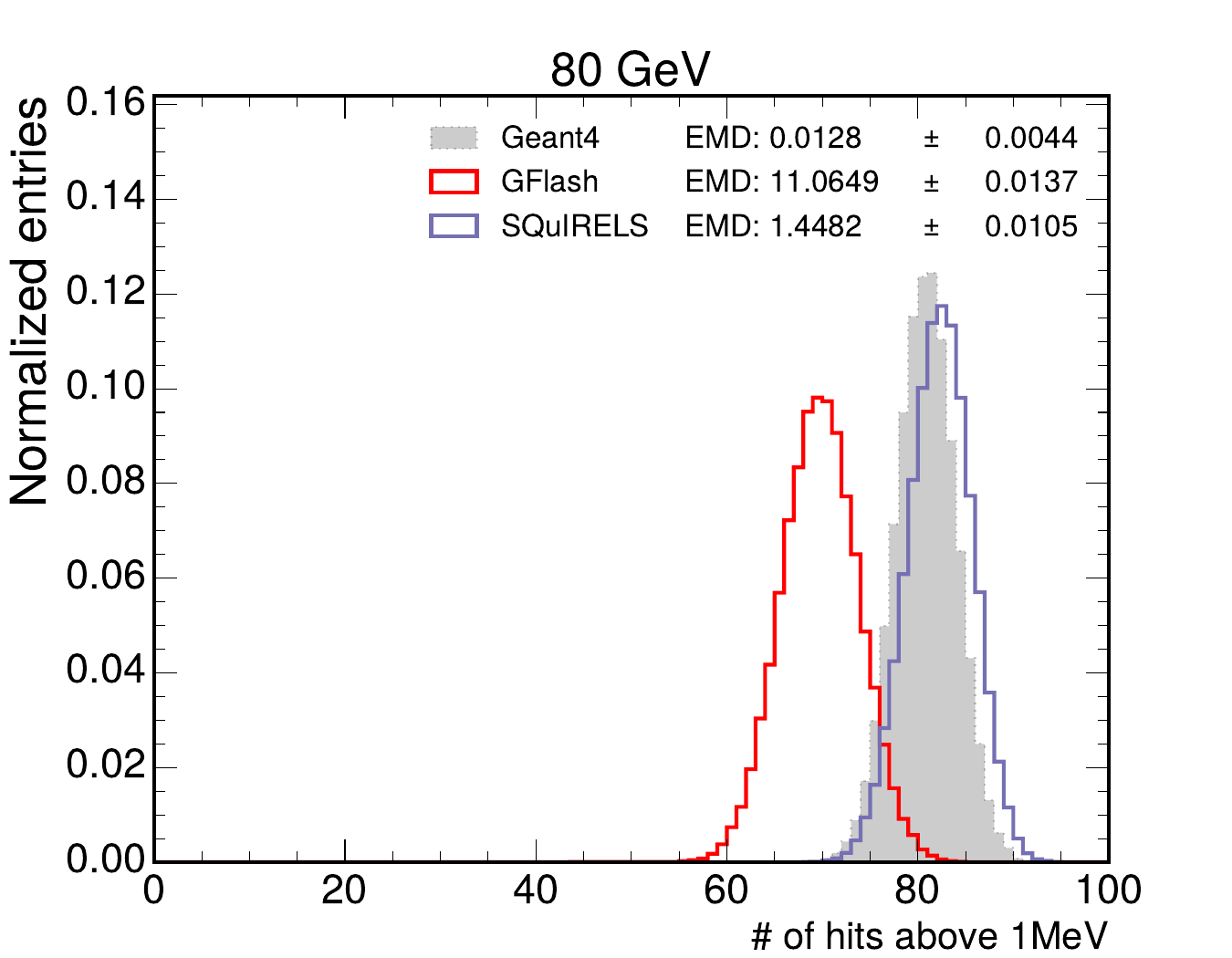}
    \includegraphics[width=0.3\textwidth]{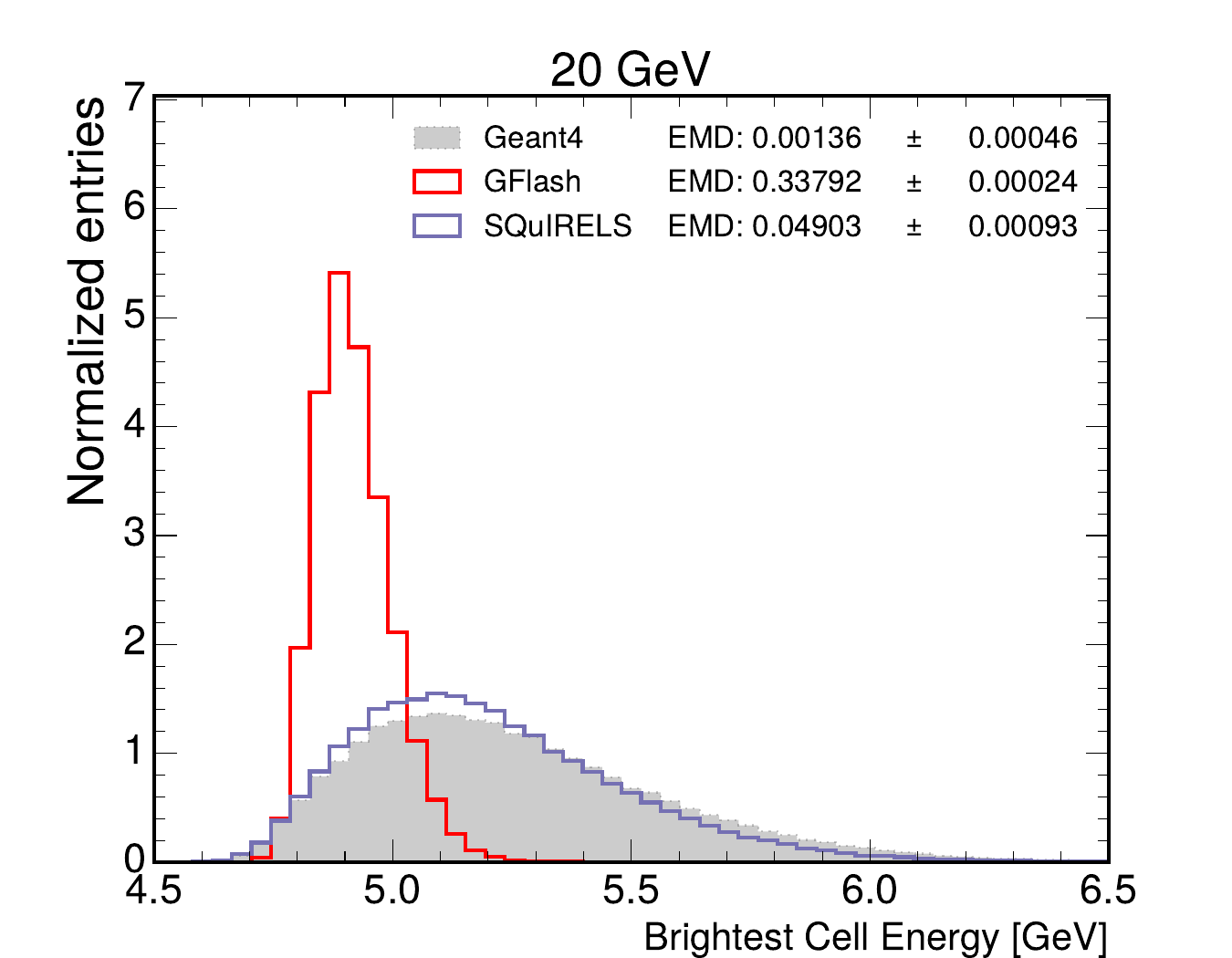}
    \includegraphics[width=0.3\textwidth]{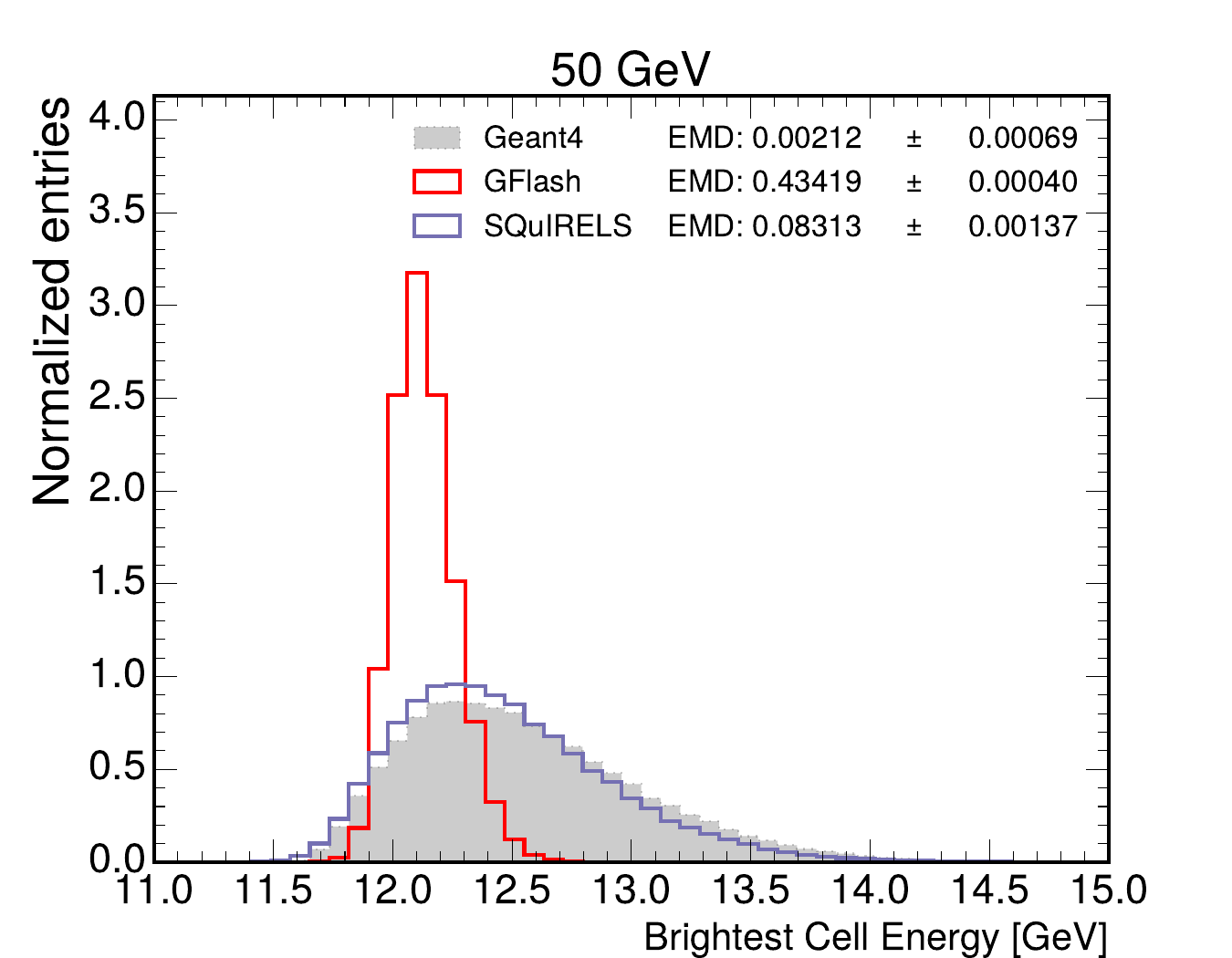}
    \includegraphics[width=0.3\textwidth]{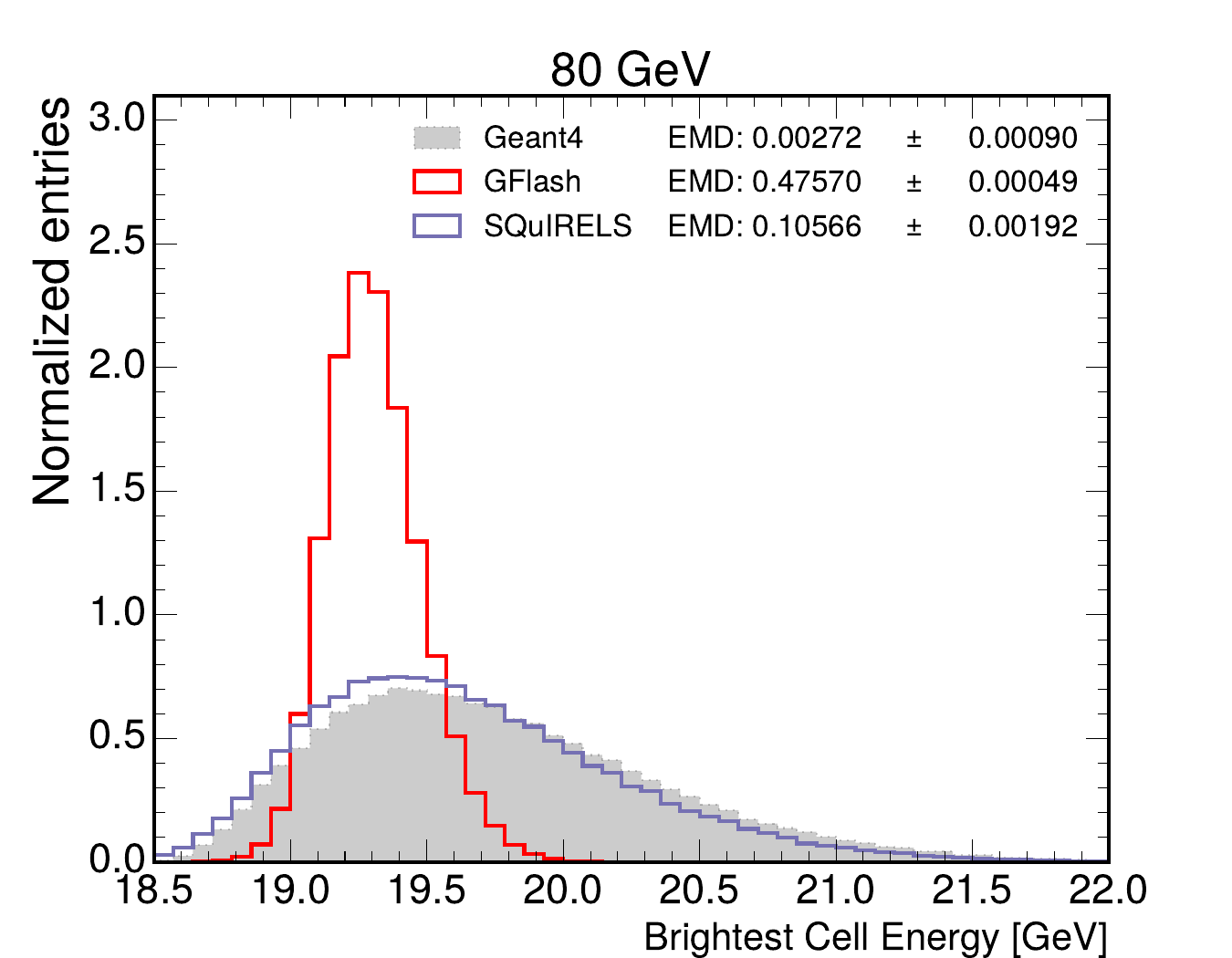}
    
\caption{Comparison between full simulation (labeled \geant), fast simulation (labeled \gf), and the Schr\"{o}dinger Bridge-refined fast shower simulation (labeled SQuIRELS). The panels show (from left to right) the total deposited energy for 20 GeV, 50 GeV, and 80 GeV electrons respectively. The top row shows the number of cells in the calorimeter with energy depositions above 1~MeV, while the lower row shows the energy deposited in the brightest cells. EMD values provide a quantitative agreement score between the reference \geant and the 3 methods, see text for more detail. }
\label{fig:SB_nhits}
\end{figure*}

The lower row of Fig.~\ref{fig:SB_esum} shows the cell energy spectra. The \geant spectra show a clear three-peak structure, with each peak corresponding to a separate region in the calorimeter. The peak in the high-energy region corresponds to the central 4 cells located around the impact point of the electrons. These cells contain the bulk of the energy deposited by the electron shower, resulting in a separate peak. Similarly, the second peak is associated with the cells directly neighboring the central 4. On average, the neighboring cell received a higher amount of energy than any surrounding ones, leading to another distinct peak. This behavior is also present for further out cells, however, their associated peaks become a lot wider and blend together into the low-energy bulk of the spectra. It can be seen that \gf approximately recreates the structure of the energy spectra, however, only with severe deviations. The spectra of the SqUIRELS-refined showers, on the other hand, match \geant very closely until around 1~MeV. While having the SqUIRELS showers agree even below 1~keV would be ideal, any realistic setting would likely discard hits below or around 1~MeV, as readout noise becomes a concern below this energy.

Next, we investigate observables derived directly from the cell energy spectra. Fig.~\ref{fig:SB_nhits} shows the number of cells with energy depositions above 1~MeV in the top row, and the energy value of the brightest pixels in the bottom row. From the number of hits figures, it can be seen that \gf consistently underestimates this observable across all shown energies. This mismatch can be traced back to the deviations in the \gf energy spectrum discussed in the previous paragraph. Further, one can see that SqUIRELS' number of hits-distributions matches up remarkably well with \geant for the 50~GeV showers, and only shows comparatively small deviations for 20 and 80~GeV. This presents a significant improvement over the \gf baseline.

\begin{figure*}[ht]
\centering
    \includegraphics[width=0.3\textwidth]{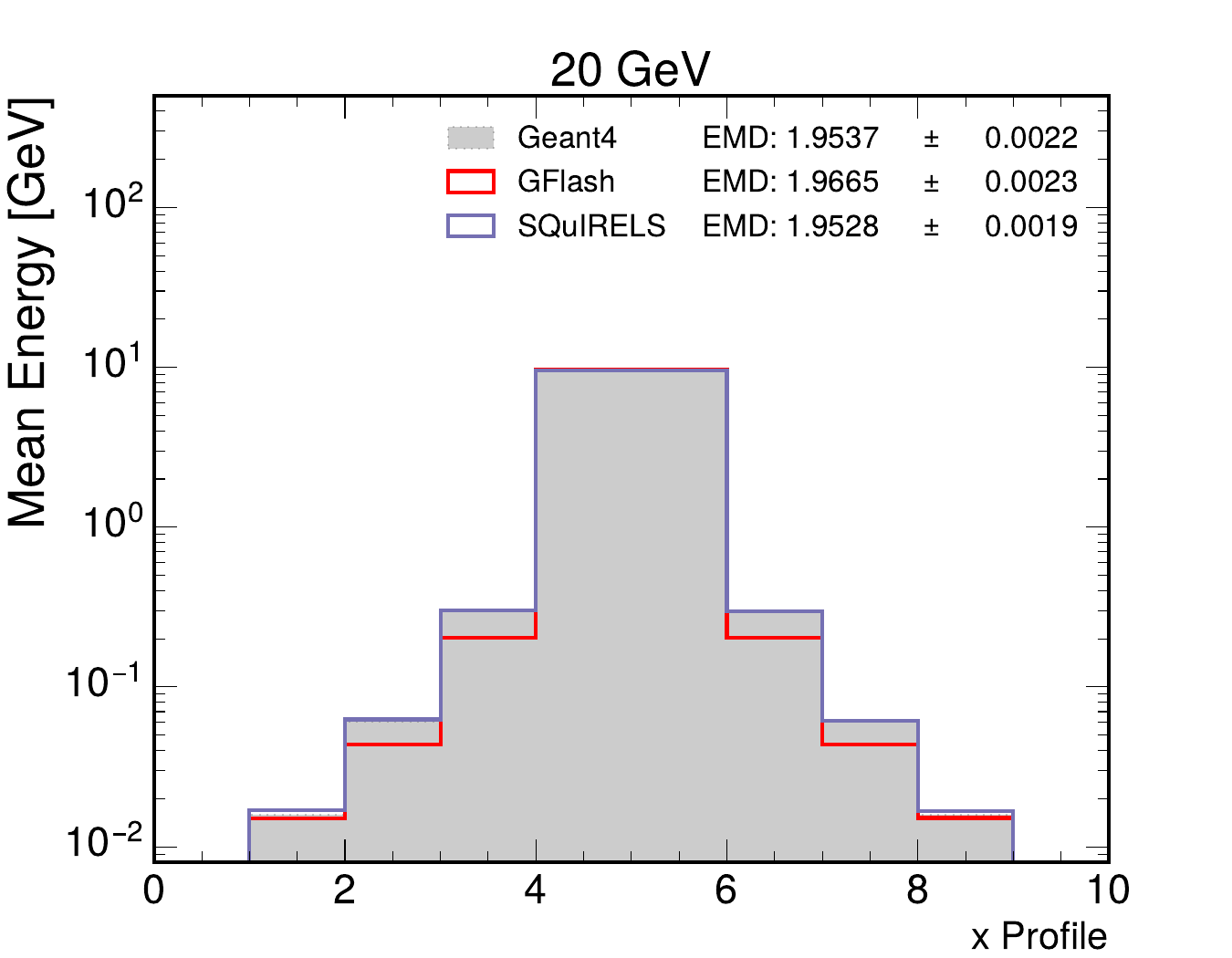}
    \includegraphics[width=0.3\textwidth]{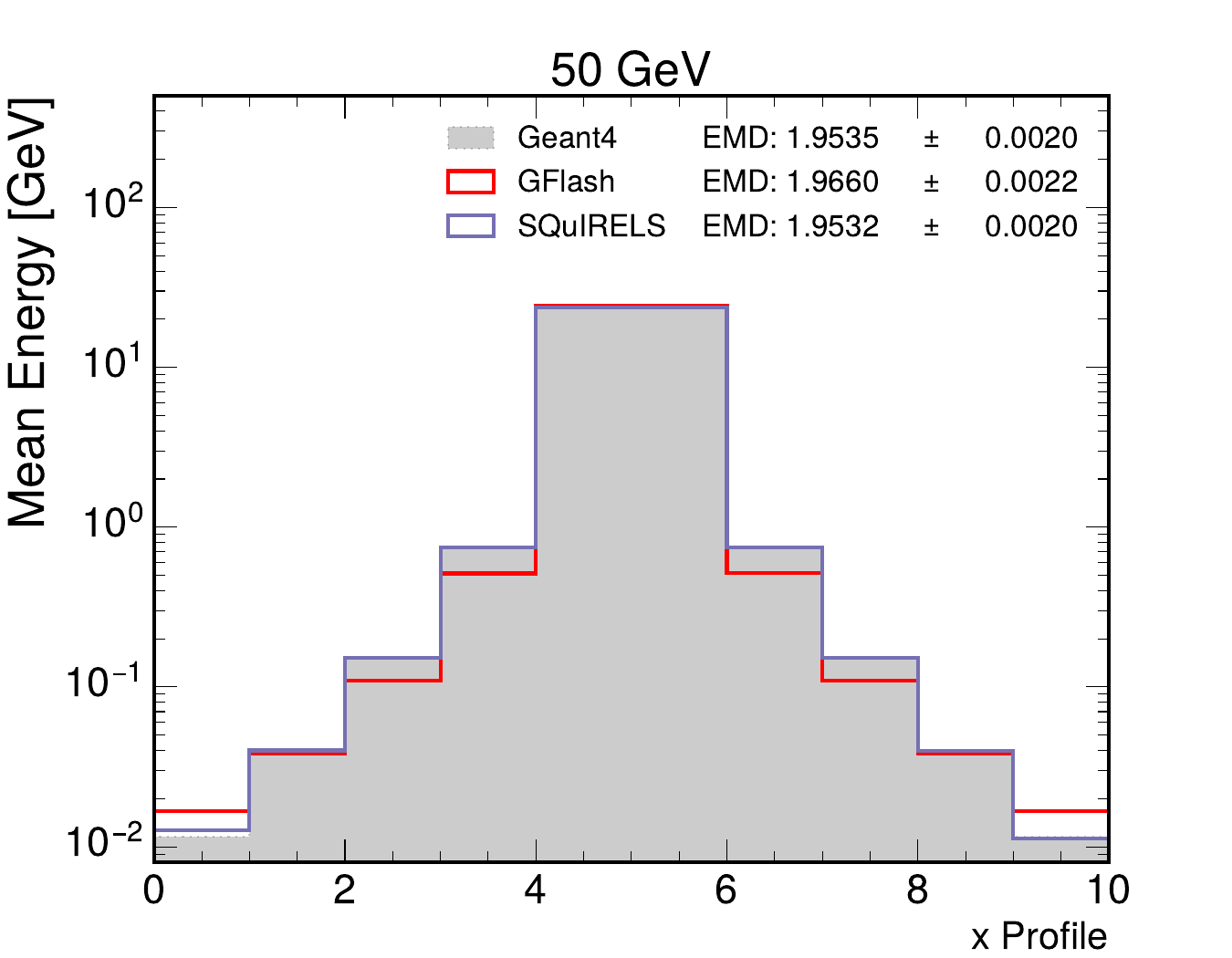}
    \includegraphics[width=0.3\textwidth]{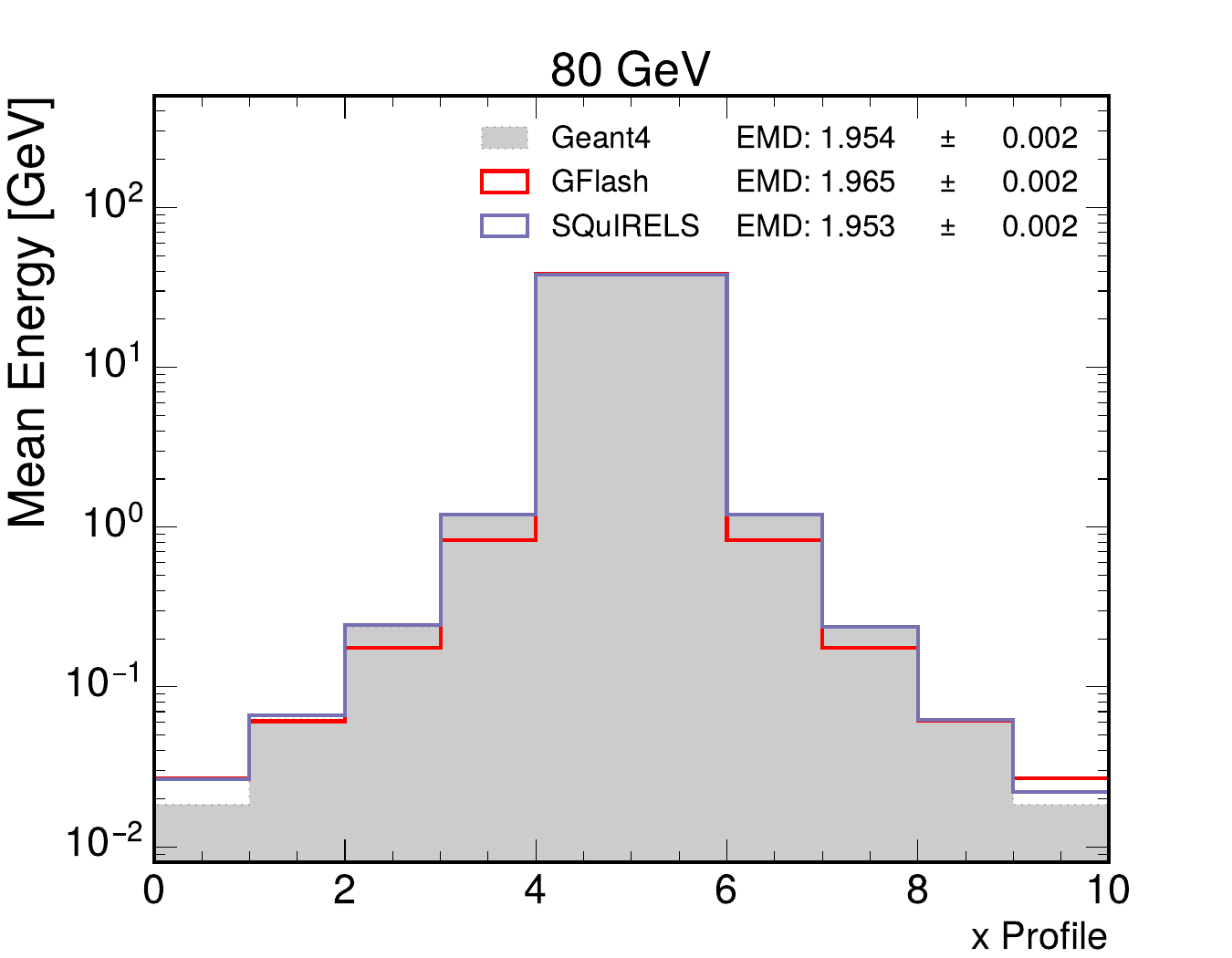}
    \includegraphics[width=0.3\textwidth]{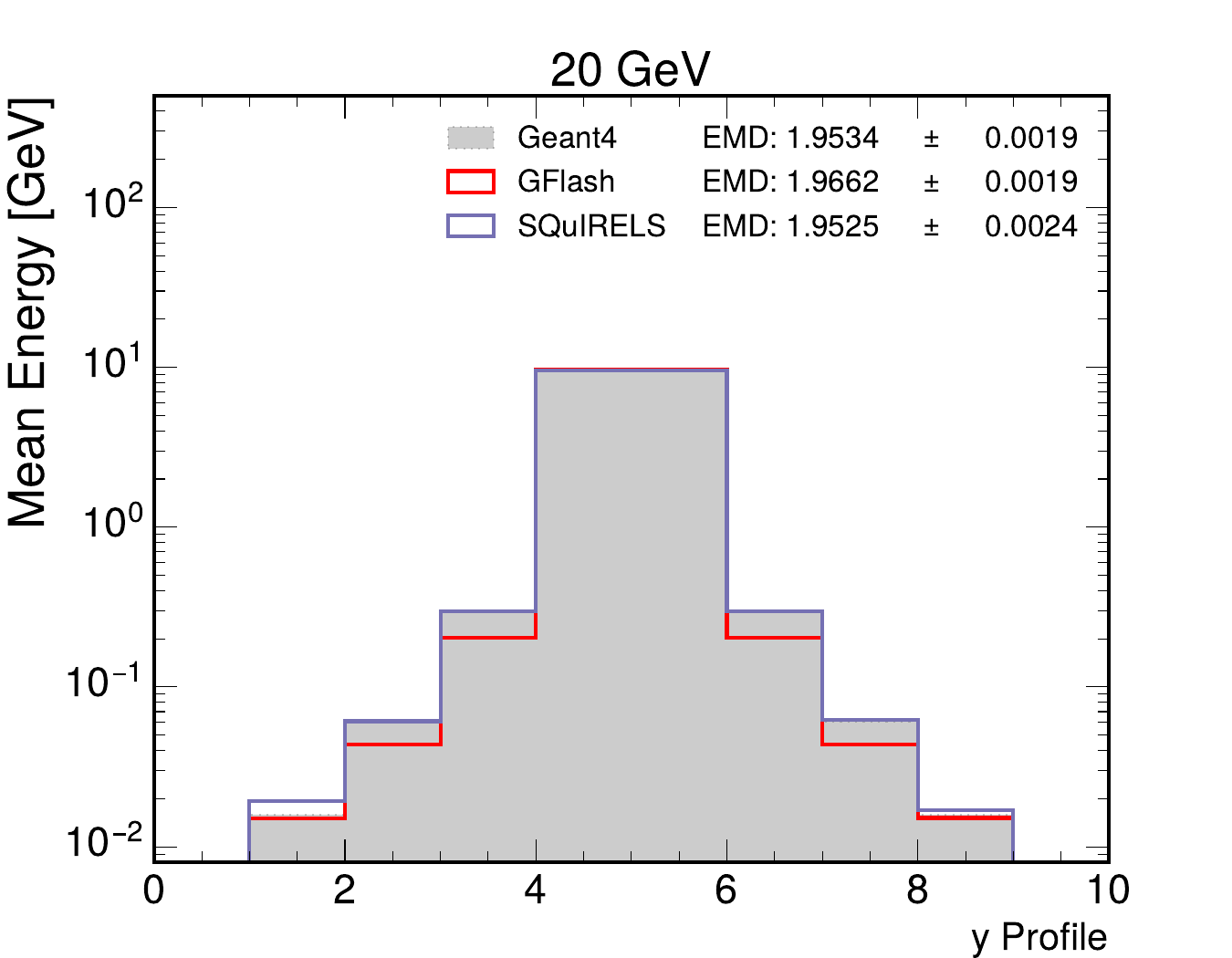}
    \includegraphics[width=0.3\textwidth]{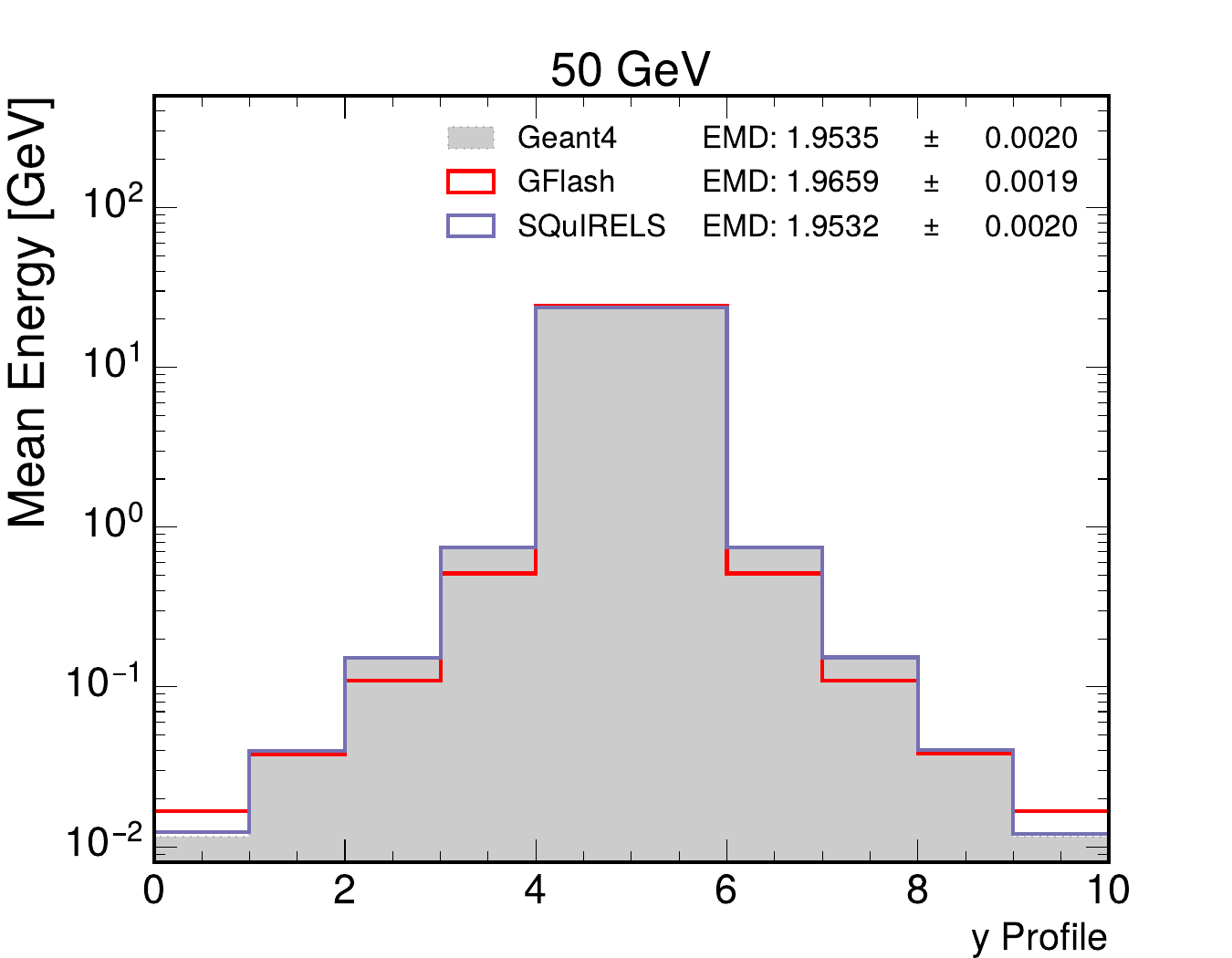}
    \includegraphics[width=0.3\textwidth]{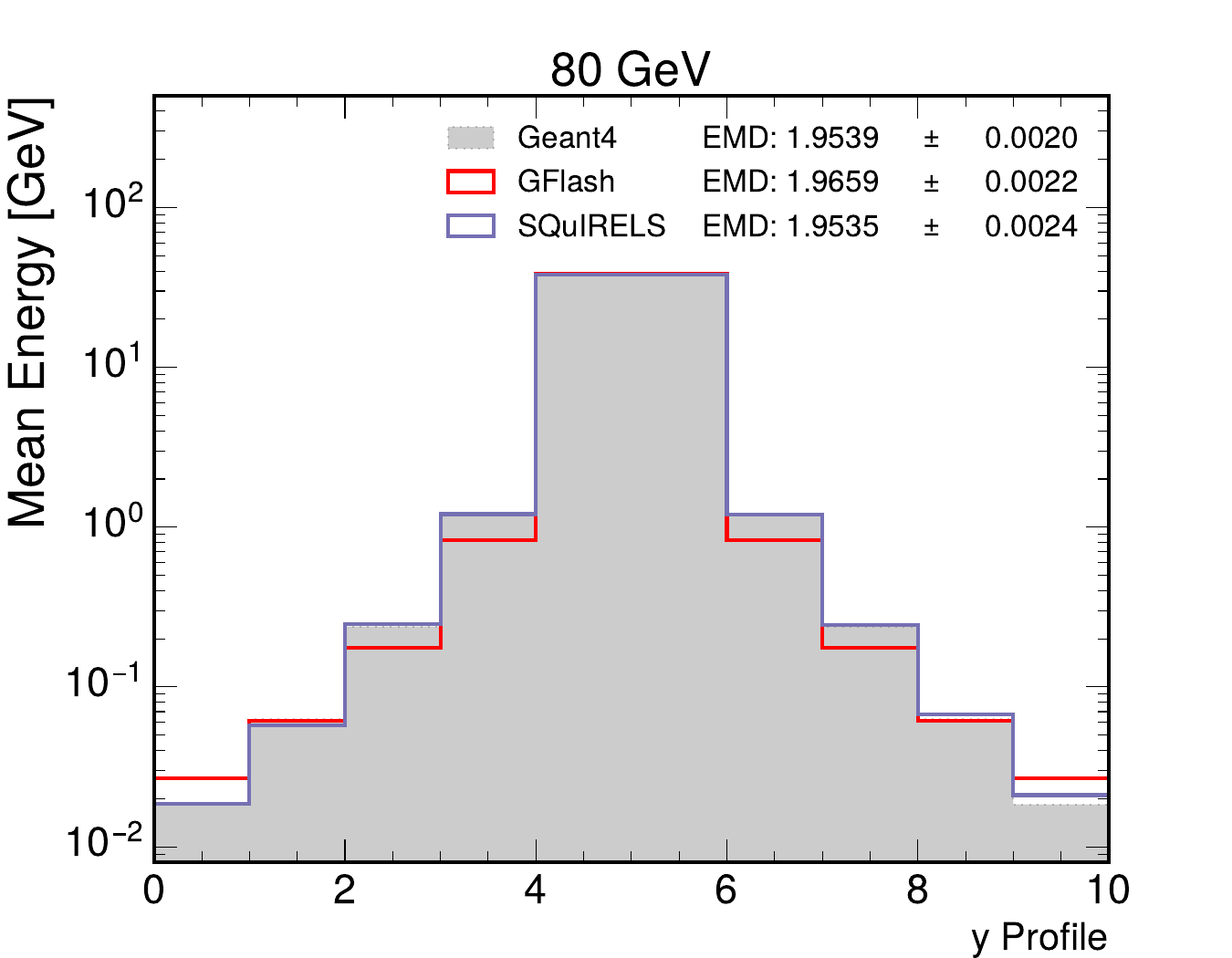}
    
\caption{Comparison between full simulation (labeled \geant), fast simulation (labeled \gf), and the Schr\"{o}dinger Bridge-refined fast shower simulation (labeled SQuIRELS). The panels show (from left to right) the total deposited energy for 20 GeV, 50 GeV, and 80 GeV electrons respectively. The top row shows the energy profile in the x direction, while the lower row shows the profile in the y direction. EMD values provide a quantitative agreement score between the reference \geant and the 3 methods, see text for more detail. }
\label{fig:SB_profiles}
\end{figure*}

\begin{table}[ht]
    \centering
	\small
    \caption{Summary of the EMD values between the 3 shown approaches and the reference \geant data sets. More details on the EMD calculation can be found in the text. A comparison between \geant and itself is included to determine the lower limit on the EMD caused by the statistical nature of the bootstrapping. }
    \label{tab:EMD}
	\begin{tabular}{l|c|c|c|c|c|cc}
        Observalbe &   \multicolumn{3}{c}{EMD} \\
          &   \geant & \gf & SQuIRELS\\
        \hline     
        $\textrm{E}_{\textrm{sum}}$ 20 GeV & 0.0003(1) & 0.1184(2) & 0.0061(2) \\
        $\textrm{E}_{\textrm{sum}}$ 50 GeV & 0.0009(2) & 0.2599(6) & 0.0147(4) \\
        $\textrm{E}_{\textrm{sum}}$ 80 GeV & 0.0016(5) & 0.3906(8) & 0.0288(8) \\        \hline     
        $\textrm{E}_{\textrm{spec}}$ 20 GeV & 0.002(1) & 0.018(2) & 0.003(2) \\
        $\textrm{E}_{\textrm{spec}}$ 50 GeV & 0.006(4) & 0.034(3) & 0.007(4) \\
        $\textrm{E}_{\textrm{spec}}$ 80 GeV & 0.009(6) & 0.051(6) & 0.011(6) \\       \hline     
        $\textrm{N}_{\textrm{hit}}$ 20 GeV & 0.015(4) & 11.37(1) & 1.32(1) \\
        $\textrm{N}_{\textrm{hit}}$ 50 GeV & 0.015(5) & 11.84(1) & 0.085(7) \\
        $\textrm{N}_{\textrm{hit}}$ 80 GeV & 0.013(4) & 11.06(1) & 1.45(1) \\       \hline     
        $\textrm{E}_{\textrm{max}}$ 20 GeV & 0.0014(5) & 0.3379(2) & 0.0490(9) \\
        $\textrm{E}_{\textrm{max}}$ 50 GeV & 0.0021(7) & 0.4342(4) & 0.083(1) \\
        $\textrm{E}_{\textrm{max}}$ 80 GeV & 0.0027(9) & 0.4757(5) & 0.106(2) \\        \hline     
        $\textrm{profile}_{\textrm{x}}$ 20 GeV & 1.954(2) & 1.967(2) & 1.953(2) \\
        $\textrm{profile}_{\textrm{x}}$ 50 GeV & 1.954(2) & 1.966(2) & 1.953(2) \\
        $\textrm{profile}_{\textrm{x}}$ 80 GeV & 1.954(2) & 1.965(2) & 1.953(2) \\       \hline
        $\textrm{profile}_{\textrm{y}}$ 20 GeV & 1.953(2) & 1.966(2) & 1.953(2) \\
        $\textrm{profile}_{\textrm{y}}$ 50 GeV & 1.954(2) & 1.966(2) & 1.953(2) \\
        $\textrm{profile}_{\textrm{y}}$ 80 GeV & 1.954(2) & 1.966(2) & 1.953(2) \\

	\end{tabular}
\end{table}

A similar trend can be seen in the brightest pixels distributions in the bottom row of Fig.~\ref{fig:SB_nhits}. Here, the \gf distribution shows large deviations compared to \geant, both in the position of the central value and the widths of the distributions. We can, again, observe a  notable improvement after applying SqUIRELS, leading to a good agreement between the brightest pixel distributions of SqUIRELS and \geant. 

Finally, we can examine the transversal profiles of the showers produced by the three methods. The top row of Fig.~\ref{fig:SB_profiles} shows the average energy profiles along the X-direction, while the bottom row shows the profiles in the Y-direction \footnote{As the detector setup is perfectly symmetric, the assignment of X and Y for the two directions orthogonal to the propagation direction of the shower is arbitray}. For both directions, one can see that compared to \geant, \gf tends to deposit more energy in the central cell and less energy in the surrounding one. As with the previously examined distributions, SqUIRELS manages to significantly improve the agreement with \geant. 

Table~\ref{tab:EMD} summarizes the EMD between the \geant baseline and \geant, \gf and SqUIRELS for all shown distributions. From the table, one can see that for all observables, SqUIRELS results in significantly smaller EMD values than \gf. Notably, the x- and y-profiles display a high baseline EMD, even between two \geant samples, however, the relative trend of SqUIRELS achieving smaller EMD values compared to \gf is still visible.

\subsection{Computational Timings}

The fundamental goal of SQuIRELS is to provide a faster simulation setup than what can be achieved using classical simulation. We, therefore, quantify the speedup gained by using \gf and SQuIRELS compared to running \geant on both CPU and GPU setups. The results of these measurements are shown in table~\ref{tab:timing}. 

For the CPU timing, 10 separate simulations of 100 events were run on an AMD EPYC 7713 64-Core Processor. All processes were constrained to a single core to ensure comparability. The GPU times were obtained by running 10 separate refinements of 100,000 events on a single Nvidia A100-SXM 40~GB GPU with a batch size of 10,000. The quoted numbers represent the average over the 10 runs, while the uncertainty is obtained from the standard deviation of the same runs. 

Table~\ref{tab:timing} shows that even on a CPU-based system, the time required for SQuIRELS to refine a shower is around $50\times$ faster than a full \geant simulation. To allow for a realistic comparison, however, one needs to account for the time needed to produce the to-be-refined \gf showers. This reduces the speedup factor of the full SQuIRELS approach to approximately $25\times$, which still presents a significant speedup over \geant. 

When run on a GPU setup, the SQuIRELS refinement is around $7700 \times$ faster than \geant, however, at this point the prerequisite \gf simulation becomes a significant bottleneck, taking up 99.4\% of the full SQuIRELS simulation time. This means SQuIRELS does benefit less from moving to GPU-based systems compared to other ML fastsim methods, but also makes SQuIRELS a natural fit for current CPU systems. 

\begin{table}[ht]
    \centering
	\small
    \caption{Comparison of the per shower computation times for \geant, \gf, and SQuIRELS. The SQuIRELS (full) timing includes the time for running the prerequisite \gf simulation. Details on the timing measurements can be found in the text.}
    \label{tab:timing}
	\begin{tabular}{l|c|c|c|c|c|cc}
        Simulator &   CPU [ms/shower] & GPU [ms/shower] \\
        \hline     
        \geant & 404.8$\pm$8.5 & / \\
        \gf & 8.5$\pm$0.4 & / \\
        SQuIRELS (refine) & 7.21$\pm$0.04 & 0.0522$\pm$0.0002 \\
        SQuIRELS (full) & 15.7$\pm$0.4 & 8.5$\pm$0.4 \\
	\end{tabular}
\end{table}

\section{Conclusion and Outlook}
\label{sec:conclusions}

We introduced SQuIRELS, a Schr\"{o}dinger bridge-based approach for refining existing fast simulation. SQuIRELS leverages the ability of a Schr\"{o}dinger bridge to map between arbitrary distributions to allow for 1-to-1 refinement without having to rely on reweighting methods. 

We benchmark SQuIRELS by refining a \gf fast calorimeter simulation of a mock detector to a full \geant-based simulation. We initially demonstrate how a Schr\"{o}dinger Bridge can transform a one-dimensional distribution by refining the energy sum distribution of the simulated showers, before moving to the refinement of the full 100-pixel showers. Our benchmark shows that SQuIRELS can significantly improve the overall quality of the showers, while still providing a significant speedup over \geant, even if run on a CPU setup.

While a cursory comparison to ab initio surrogate modeling shows that the refinement approach exhibits no notable benefits compared to the ab initio approach on the simplified calorimeter data shown in this work, there exists a wide range of potential applications that may be able to better leverage the features of the SQuIRELS refinement. 

For now, this work serves as a proof of concept for using Schr\"{o}dinger Bridges for the purpose of 1-to-1 calorimeter simulation refinement. 
Further application to more complex data sets will likely present further computational and implementational challenges.  
Furthermore, SQuIRELS imposes no constraints on possible network architectures, making it a great candidate for the application to high-dimensional or even point cloud-based data sets in the future.

\section*{Code Availability}

The code for this paper can be found at \url{https://github.com/SaschaDief/SB_refinement}. The data is available from Zenodo~\url{https://doi.org/10.5281/zenodo.8275193}.

\section*{Acknowledgments}
SD, VM, and BN are supported by the U.S. Department of Energy (DOE), Office of Science under contract DE-AC02-05CH11231. This research used resources of the National Energy Research Scientific Computing Center, a DOE Office of Science User Facility supported by the Office of Science of the U.S. Department of Energy under Contract No. DE-AC02-05CH11231 using NERSC award HEP-ERCAP0021099.  

\bibliography{HEPML,other}
\bibliographystyle{apsrev4-1}

\clearpage
\appendix

\end{document}